***Emergent Ferroelectric Nematic and Heliconical Ferroelectric Nematic States***
***in an Achiral "Straight" Polar Rod Mesogen***

Hiroya Nishikawa*, Daichi Okada, Dennis Kwaria, Atsuko Nihonyanagi, Motonobu Kuwayama, Manabu Hoshino and Fumito Araoka*

RIKEN Center for Emergent Matter Science (CEMS), 2-1 Hirosawa, Wako, Saitama 351-0198, Japan

Graduate School of Medicine, and General Medical Education and Research Center, Teikyo University, 2-11-1, Kaga, Itabashi-ku, Tokyo 173-8605, Japan

Present address (Dr. Daichi Okada): Faculty of Electrical Engineering and Electronics, Kyoto Institute of Technology, Matsugasaki, Sakyo-ku, Kyoto, 606-8585, Japan

E-mail: hiroya.nishikawa@riken.jp; fumito.araoka@riken.jp

**Abstract**

Ferroelectric nematic liquid crystals (NFLCs) are distinguished by their remarkable polarization characteristics and diverse physical phenomena, sparking significant interest and excitement within the scientific community. To date, over 150 $N_F$LC molecules have been developed; however, there are no reports regarding straight linear polar molecules with a parallel alignment of the permanent dipole moment and the molecular axis. The straight polar mesogen **nBOE** exhibited an enantiotropic $N_F$ phase with a wide temperature window (up to 100 K) despite having a longer alkyl chain (up to n = 6) than the critical alkyl chain length of conventional models. Interestingly, **nBOE** with a medium-length alkyl chain displayed an exotic phase sequence of $N_F$–$^{HC}N_F$–$SmC_F$ during the elimination of positional displacement among adjacent molecules. Furthermore, we demonstrate the reflective color modulation of the $^{HC}N_F$LC over the entire VIS-NIR spectral regime by ultralow $E$-field (up to 0.14 V $\mu m^{-1}$).

# 1. Introduction

Giant fluid ferroelectricity emerges in a new class of matter states called the ferroelectric nematic ($N_F$) phase,[1–3] which is described by a long-range polar orientational order. Thus, the $N_F$ phase has a global $C_{\infty v}$ symmetry since the macroscopic polarization aligns along the director (Figure 1c). Usually, the $N_F$ phase can be formed by a liquid crystalline (LC) rod-like molecules with high dipole moment (> 9 Debye),[4] exhibiting giant polarization behavior, i.e., as apparent dielectric permittivity (< ~10k),[5] polarization density (> ~4 μC cm$^{-2}$),[5] and NLO coefficient (< ~10 pm V$^{-1}$) [5] and unique physical properties such as topology,[6,7] instability,[8,9] fiber,[10,11] thermomotor[12] as well as superscreening[13]. These outstanding characteristics have led to exponential growth in state-of-the-art ferroelectric research. In particular, research dedicated to a deeper understanding of the relationship between the molecular structure and $N_F$ phases is flourishing, with over 150 types of $N_F$LC molecules developed to date.[4] The structure of $N_F$LC molecules is highly delicate and difficult to tailor without inspiring the backbone of the archetypal $N_F$LC molecules (i.e., DIO,[1] RM734,[2] and UUQU-4-N[14]).

However, some studies have provided valuable insights into molecular design. Mandle et al. [15] and Chen et al. [3] independently reported that a strong dipole–dipole interaction along the director triggers the emergence of the $N_F$ phase with the aid of molecular dynamics simulations. Madhusudana introduced a model describing a molecule (a rod entity) in the $N_F$ phase by a longitudinal surface charge density wave. [16] In this model, the adjacent molecules orient in a syn-polar fashion with a molecular offset such that the amplitude of the charge density waves is minimized (thus reducing the electrostatic energy), further indicating that this polar arrangement emerges in a high-density state. Notably, the model with an alternating charge distribution along the molecular axis was consistent with $N_F$LC molecules. Cruickshank et al. obtained good feedback on the validity of Madhusudana within RM734 families.[17–19] Nacke et al. discussed Mandle and Madhusudana's model in the $N_F$ phase for AUUQU-2-N using synchrotron-based X-ray diffraction (XRD) studies.[20] A smectic C-type long-range correlation is suggested over the entire temperature range of the $N_F$ phase for AUUQU-2-N. In their proposed model, the molecules packed in a layer are slightly displaced from each other, forming a polar structure (a similar model reported in our previous study[21]). More recently, Marchenko et al. directly visualized synpolar molecular ordering with a molecular offset in a monolayer on an Au(111) surface.[22]

Much effort has been made to develop $N_F$LCs, but the dipole moment of $N_F$LC generics (>150 types) forms an angle between 10° and 25° with the long axis (Figure 1a). This property is attributable to the molecular structure consisting of a polar linker (e.g., COO, $CF_2$O) and a polar end unit (e.g., 1,3-dioxane, ester[23]), which both have dipole moments that deviate from the molecular axis. Furthermore, the common characteristics of the $N_F$ phase in conventional models include being thermodynamically metastable and significantly destabilized (or vanishing) as the length of the alkyl chain increases owing to the reduction of the on-axis dipole-dipole interaction (Figure 1c). It is of paramount importance to develop a molecular design that can overcome the universally observed drawbacks.

In this paper, we present a new model, **nBOE** (n = 1–8), with a hard-rod polar molecule motif, in which the direction of the dipole moment and the molecular axis are in an approximately perfect parallel alignment (Figure 1d). Notably, we found that the **nBOE** variants exhibited not only an enantiotropic $N_F$ phase with a wide temperature window but also an exotic ferroelectric fluid phase transition via $N_F$, heliconical $N_F$ ($^{HC}N_F$), and ferroelectric smectic C (SmC_F) phases (Figure 1e and 1f). Recently, new polar helical phases ($N_{TBF}$[24] and $SmC_p^H$[25]) have been found in achiral molecules that combine the cores of DIO, RM734, and UUQU. However, we have opened the way for exploring novel helielectric phases in ferroelectric fluid libraries of rigid rod mesogen, along with new helielectric phases in even just "straight" polar rods.

## 2. Results and Discussion

### 2.1. Molecular Structure of BOE series.

The chemical structures of the **BOE** series are shown in Figure 1d. Compared to the common motifs (RM734, DIO, and UUQU-4-N), the structure of **BOE** does not bear a flexible linker such as ester (COO) or difluoromethoxy ($CF_2O$) units; instead, it incorporates a rigid diphenyl alkyne unit, well-known as tolan. Furthermore, to increase the whole dipole moment ($\mu$) and to prevent deviation of the dipole from the molecular axis, we end-capped the tolan unit using a bicycloorthoester unit, designing the **BOE** molecules. In such a straight, hard rod molecule, the dipole moment may direct along the long axis of the molecule. As expected, the **nBOE** series (n = 1–8) showed a small $\beta$ angle ranging between 0.25° and 4.7° owing to the hard-rod molecular design. The optimized structures of **nBOE** obtained using DFT calculations are displayed in Figure S13a, Supporting Information. The calculated $\mu$ and $\beta$ as a function of the number of carbon atoms (n) for **nBOE** are shown in Figure S13b and S13c and Table S1, Supporting Information. **1BOE–3BOE** (short-alkyl chain group) showed a negligibly small $\beta$ (<0.3°), indicating the dipole moment and the molecular axis are perfectly parallelly aligned. With increasing alkyl chain length, $\beta$ also increased to between 1° and 3° for **4BOE–6BOE** (medium-alkyl chain group). **7BOE** and **8BOE** (long-alkyl chain group) still showed small angles of $\beta$ = 3.4° and 4.7° respectively. Lengthening the alkyl chain slightly increased the $\mu$ value, marking ca. 15 Debye for all the **BOE** series. Figure 1e and Figure 2a–2d show the phase transition behaviors and DSC curves of **nBOE**, respectively. For short-alkyl chain groups (n = 1–3), the $N_F$ phase appeared at the melting point of the pristine crystal, and the $N_F$ phase emerged again from the upper phase [isotropic liquid (IL) or nematic (N)], indicating enantiotropy of the $N_F$ phase in these compounds. The $N_F$ phase is usually characterized by $2\pi$-twist walls (Figure 2e).[ 26] Medium-alkyl chain groups (n = 4–6) were also found to be enantiotropic $N_F$LCs. On the other hand, the $N_F$ phase was excluded for longer alkyl chain groups (n = 7–8); instead, apolar phases (N′ and SmA) appeared. With increasing molecular length, the N phases appeared as a highest temperature mesophase. In addition, in the case of n=5–6, the $M_{AF}$ phase with the typical zig-zag texture and very small enthalpy ($10^{-3}$–$10^{-2}$ kJ mol$^{-1}$ order, Figure 2c and 2f) was observed. Notably, **nBOE** (n = 1–5) exhibited a broadened temperature window for the $N_F$ phase (~50–100 K) during heating. Interestingly, **nBOE** induces the $N_F$ phase even with long alkyl chains, which significantly differs from conventional $N_F$LCs. For instance, the $N_F$ phases exhibited in RM734 and DIO (hereafter referred to as **nRM** and **nDIO**, respectively) are thermodynamically metastable (see Figure S14, Supporting Information). Irrespective of the alkyl chain length, **nRM** exhibited a monotropic $N_F$ phase, which barely emerged for n = 2 (**2RM**). By contrast, **nDIO** with short alkyl chains (n = 1–2) is an enantiotropic $N_F$ phase, but the temperature windows are quite narrow (<7 K) upon heating. The metastable $N_F$ phase was observed narrowly within **4DIO**. In sharp contrast, the $N_F$ phase remained alive in **nBOE** up to n = 6. Unexpectedly, two extra ferroelectric phases, heliconical ferroelectric nematic ($^{HC}N_F$) and ferroelectric smectic C (SmC$_F$), emerged below the $N_F$ regime for **nBOE** (n = 4–6) (Figure 1e).

## 2.2. Phase Transition Behavior and Unique Polarized Optical Textures.

The common DSC features of **nBOE** (n = 4–6) within the $N_F$–$^{HC}N_F$–SmC$_F$ regime were a baseline step ($N_F$–$^{HC}N_F$) and a distinct exothermal peak ($^{HC}N_F$–SmC$_F$), as shown in Figure 2d and Figure. S15–S17, Supporting Information. Figure 2g displays the unique texture change through the unique cascade phase changes of **5BOE**. During cooling from the $N_F$ phase, the reddish texture gradually appeared at 145 °C (panel (v) in Figure 2d and 2g) in a parallel-rubbed cell. Further cooling led to a gradual color change (blue shift) in the POM texture. However, in this state, vivid blue reflection can be observed when viewed by naked eyes, indicating that a specific helical structure exists in the $^{HC}N_F$ phase (complete images are shown in Figure S18a, Supporting Information). Notably, this texture of the $^{HC}N_F$ phase differed from the planar texture with defects (a.k.a., oily streaks) of the helicoidal ferroelectric nematic, $^{HD}N_F$ (o.k.a., chiral ferroelectric nematic, $N_F^*$) phase in the rubbed cell (Figure S19, Supporting Information).[27,28] At ~125 °C, a drastic change in texture with strong light scattering occurred (panel (xi) in Figure 2d and 2g). The textural difference between $^{HC}N_F$ and $SmC_F$ phases was more remarkable in an antiparallel-rubbed cell (panels (xi′) and (xii′) in Figure 2g and Figure S18b, Supporting Information). In the $^{HC}N_F$ phase, a striped texture tilted at a certain angle relative to the rubbing direction appeared. Similarly, a striped texture emerged in the $SmC_F$ phase; however, unlike in the $^{HC}N_F$ phase, the orientation of the stripes was parallel to the rubbing direction. For the $^{HC}N_F$ phase, similar behavior is usually observed in twist-bend nematic ($N_{TB}$) materials, including a second-order-like $N–N_{TB}$ transition and a striped texture.[29] The details of the $^{HC}N_F$ and $SmC_F$ phases are discussed in Section 2.5 and Supporting Note 2, Supporting Information, respectively.

## 2.3. Ferroelectric Behavior of the $N_F$, $^{HC}N_F$, and $SmC_F$ Phases.

To evaluate the ferroelectricity of the three polar phases ($N_F$, $^{HC}N_F$, and $SmC_F$) for **nBOE**, we performed dielectric relaxation (DR), polarization reversal current (or $P–E$ hysteresis), and second harmonic generation (SHG) measurements. Figure 3a–d shows the results of the DR studies for **3BOE** and **5BOE**. For both cases, the giant dielectric permittivity ($ε′$) of ~6k–8k (**3BOE**) and ~6.8k (**5BOE**) was observed in the $N_F$ regime, corresponding to its magnitude for typical $N_F$LCs. For **5BOE**, with decreasing temperature, the $ε′$ value in the $N_F$ phase experienced a minor increase, followed by a gradual decrease (down to ~6.5k) upon entering the $^{HC}N_F$ regime. The $ε′$ value dropped to ~6.0k at the $^{HC}N_F–SmC_F$ phase transition temperature, continuing to reduce $ε′$ toward the low-temperature side (Figure 3d). The relaxation frequency was slightly increased via the $N_F → ^{HC}N_F → SmC_F$ phase transition (Figure 3c, Figure S20, Supporting Information).

For the polarization reversal current measurement, no current peak was observed in the N phase for **3BOE**, whereas a distinct peak appeared in the $N_F$ phase for **3BOE** and **5BOE** (Figures 3e and 3f). As shown in the corresponding $P$–$E$ hystereses (insets in Figure 3e and 3f), a parallelogram hysteresis loop, commonly observed in ferroelectrics, was obtained for the $N_F$, $^{HC}N_F$, and $SmC_F$ phases. The temperature dependence of $P$ is shown in Figure 3g (**3BOE**) and 3i (**5BOE**). The average spontaneous polarization ($P_{ave}$) in the $N_F$ regime was nearly the same ~6.1 μC cm$^{-2}$ for **3BOE** and **5BOE**. For **5BOE**, in the $^{HC}N_F$ and $SmC_F$ regime, large $P_{ave}$ values of ~6.5 and ~6.8 μC cm$^{-2}$ were observed, respectively. As a first approximation, the polar ordering, <P1>, in the $N_F$ phase for **nBOE** (n = 1–6) was estimated using the equation: [30]

$$<P1> \sim P_{av}/P_{max} \tag{1}$$

$$P_{max} = \mu(\rho/M)N_A = 200\mu(\rho/M) \tag{2}$$

, where <P1>, $P_{max}$, $\mu$, $M$, and $N_A$ are the polar order parameter, maximum electron polarization, permanent dipole moment, molar mass, and Avogadro constant, respectively. The parameter $P_{max}$ implies that all the molecular dipoles are aligned perfectly in one direction along **n**). The obtained <P1> values are listed in Table S2, Supporting Information. Although the BOE series with a relatively long alkyl chain showed a relatively high <P1> ~0.8, **1BOE** showed the lowest <P1> ~0.6. This result suggests a relatively long alkyl chain offers a suitable molecular alignment for favorable dipole–dipole interactions. Notably, the three phases exhibited extremely small coercive $E$-field ($E_c$) of $10^{-3}$, $10^{-2}$, and $10^{-1}$ kV/cm$^{-1}$ order in the $N_F$, $^{HC}N_F$, and $SmC_F$ phases, respectively (Figure 3h and 3j). Moreover, the $^{HC}N_F$ phase inherited its unique property, with its $E_c$ being one order of magnitude lower than that of a similarly reported helielectric ($N_{TBF}$) phase.[24] Thus, these results indicate that the Tolan-bearing polar molecule can dramatically reduce the $E_c$ of various ferroelectric fluids.

Strong SHG activities in the $N_F$, $^{HC}N_F$, and $SmC_F$ phases were observed in **3BOE** and **5BOE** (further details provided in Supporting Note 2, Supporting Information). Therefore, the DR, $P$–$E$ hysteresis, and SHG studies demonstrate that the $N_F$, $^{HC}N_F$, and $SmC_F$ phases exhibit remarkable ferroelectric behavior. Similar behavior for DR, $P$–$E$, and SHG properties were observed for **4BOE** and **6BOE** (Figures S21 and S22, Supporting Information). Figure S23 (Supporting Information) shows the complete data on the polarization reversal current. By combining these studies, three mesophases (N, N′, and SmA) exhibited in **7BOE** and **8BOE** (long-alkyl chain groups) were found to be paraelectric LC phases (see Supporting Note 3, Supporting Information).

## 2.4. Structure Characterization based on X-ray Diffraction Analysis

XRD measurements were performed to characterize the LC structures, particularly the polar ordering for **nBOE** (n = 1–6). Figure S24 (Supporting Information) shows the 1D XRD patterns of **2BOE** and **3BOE**. Within the $N_F$ range, the $d$-spacing of the relatively sharp diffraction peak at small angles is in good agreement with the molecular length, indicating the presence of a short-range nematic order. Figure 4a–d shows the results of the XRD studies of **6BOE**. Figure 4a shows the two-dimensional (2D) X-ray diffractograms without a magnetic field in various phases. Irrespective of LC phase type, three distinct peaks were observed, for instant (e.g. $q =$ 2.77, 1.26, 0.48 Å$^{-1}$ in the N phase). The corresponding 1D XRD profiles are shown in Figure 4b. With decreasing temperature, the position of peak (iii) shifted to the large $q$ side owing to face-to-face molecular stacking, suggesting that the stacking distance changed. By contrast, the full width at half maximum (FWHM) of the various phases (N, $M_{AF}$, $N_F$, $^{HC}N_F$, and $SmC_F$) significantly decreased upon cooling. In particular, the strong intensity and narrow FWHM of the $SmC_F$ phase indicated that a long-range positional order was generated. The position of the primary peak varies during the LC phase (Figure 4c). Figure 4d shows the $d$-spacing as a function of temperature. Within the phase transition between N and $N_F$ across the $M_{AF}$ phases, $d$ sequentially decreases and reaches a constant value in the $N_F$ phase. Noteworthily, further cooling increased the $d$ value beyond the $N_F$–$^{HC}N_F$ phase transition until the $d$ value was saturated in the $SmC_F$ phase. Similar behaviors were observed for **4BOE** and **5BOE** (Figure S25 and S26, Supporting Information). The $d$-spacing in the $SmC_F$ phase for **4BOE** and **5BOE** was nearly according to the molecular length ($L_m$), whereas **6BOE** exhibited a slightly smaller $d$ value ($d < L_m$), which may have been due to the folded alkyl chain. Assuming that $d = L_m$ in the $SmC_F$ phase, dimerization with molecular displacement may occur in the $^{HC}N_F$ and $N_F$ phases. Indeed, the single crystal XRD (SC-XRD) results for **6BOE** indicate the presence of a cluster with a molecular offset (Figure 4e and Figure S27, Supporting Information). In the cluster, the offset structure is generated via multiple interactions such as hydrogen bonding, fluorine/$\pi$, and CN/$\pi$ (Figure 4f).

Figure 4g–j shows the changes in the Fourier transform infrared (FTIR) absorbance spectra related to specific stretching vibrations (see also Supporting Note 4, Supporting Information). With decreasing temperature, the peak position due to the stretching vibrations of $C–O–C_{BOE}$, $Ar^I$–F, $Ar^{II}$, and $C\equiv N$ continuously shifted to high wavenumber. By contrast, the vibrational peak of the core mesogen ($A^I+A^{II}+A^{III}$) shifted to a lower wavenumber. By combining the XRD and FTIR data, we propose the following model. (i) In the $N_F$ phase, multiple head-and-tail and side-by-side interactions stabilize the polar nematic arrangement. (ii) The side-by-side interactions become significant and induce the offset, leading to the emergence of the $^{HC}N_F$ phase. (iii) Further increase of the side-by-side interactions lead to a predominance of face-to-face communication, thereby transitioning to the $SmC_F$ phase but with an eliminated tilt angle. For (i), we believe that still the side-by-side interactions effectively stabilizes the $N_F$ phase, suggesting that the $N_F$ phase remains alive even in **6BOE** with a long alkyl chain. For (ii), the **nBOE** ($n = 4,5$) replaced the –CN group with either –F or –$NO_2$ groups (i.e., **nBOE-F** and **nBOE-NO₂**) did not induce the $^{HC}N_F$ phase in either case (Supporting Note 5, Supporting Information). This result suggests that specific interactions via the –CN group are crucial for inducing the $^{HC}N_F$ phase.

## 2.5. Unique Characterization of the $^{HC}N_F$ phase

As shown in Section 2.2, the $^{HC}N_F$ phase showed a reflective color, suggesting the presence of a helical structure, but it may be distinct from the typical $^{HD}N_F$ phase owing to its unusual POM texture. One possibility is that the molecules were oblique to the helical axis in the $^{HC}N_F$ phase, resulting in an oblique helicoidal or heliconical structure. When **nBOE** (n = 4–6) was injected into the bare glass sandwich cell, the reflective color was visible to the naked eye. In the case of **5BOE**, the alignment was more uniform than those of the others; therefore, unless otherwise noted, the investigation regarding the helical structure of the $^{HC}N_F$ phase was performed using **5BOE**. A relatively uniform texture is observed in the POM image of the bare glass cell; however, no oily streak texture is observed (Figure S19, Supporting Information). When the cell was observed straight on, an orange reflective color was observed, whereas a blue shift in color became apparent from an oblique angle (Figure 5a). Figure 5b shows the reflection spectra as a function of the cell rotation (oblique) angle. When the cell rotated counterclockwise, a blue shift from the red reflective color was observed. Similarly, the clockwise rotation showed a blue shift in the structural color (Figure S28a, Supporting Information). Therefore, despite the rotation direction, the constant peak displacement indicated that the helical axis of the $^{HC}N_F$ phase stands normal to the cell plane (Figure 5a). Figure 5c, 5d, and Figure S28b (Supporting Information) show the temperature-dependent spectral changes in the bare glass cell. Interestingly, unique spectral changes occurred sequentially in the $^{HC}N_F$ regime: (i) a blue shift to ~680 nm, (ii) a stationary shift at 680 nm (Figure 5d), and (iii) a redshift over 900 nm (outside the measurement range). The change in birefringence within the $^{HC}N_F$ regime was constant (Figure 5e, Figure S29, Supporting Information), coinciding with the result for (ii).

Next, we investigated the electric field (*E*-field) response of the helical structure in the $^{HC}N_F$ phase. In the planar state of the cholesteric LCs, the strong *E*-field (several V μm$^{-1}$) along the helical axis allows homeotropic alignment of LC because the elastic free energy is dominated by the electric free energy collapsing the helical structure. Similarly, a vertical direct current $E$-field was applied to the $^{HC}N_F$ phase in the ITO-coated glass cell. With a small $E$-field of 0.25 V μm$^{-1}$, the birefringence nearly disappeared (Figure 5g), suggesting the homeotropically aligned polar heliconical structure along the $E$-field. Notably, by slightly decrossing the polarizers to the scarcely observed domain boundaries, the contrast between adjacent domains was reversed (Figure 5g and 5i). This result indicated that heterochiral structures were present in each domain across the domain boundary (Figure 5j). As the intensity of the $E$-field progressively increased, the spectral width narrowed, accompanied by a simultaneous shift of its position toward the shorter-wavelength side, confirming the selective reflective colors over a wide range of wavelengths (Figure 5k and 5l). This color-change mechanism in the $^{HC}N_F$ phase is probably identical to that observed in heliconical nematics.[31] Thus, as shown in the model (Figure 5m), in the heliconical nematic, the director is tilted with some angle $\theta < \pi/2$ with respect to the helical axis. By applying the $E$-field along the axis, the director is reoriented with decreasing $\theta$, changing the helical pitch ($P$) without reorienting the helical axis. Typically, a heliconical nematic can be created by blending a chiral dopant to induce chirality in the host LCs (twist-bend nematic and typical nematic). However, it is noteworthy that the $^{HC}N_F$ phase is the helielectric version of the heliconical nematic phase so that its characteristics differ dramatically from those of the heliconical nematics: (i) the $^{HC}N_F$ phase is generated spontaneously because of the coupling between polar and chiral symmetry breaking, and (ii) the ultralow $E$-field-driven multicolor modulation is due to the coupling of polarization and voltage. The exceptional $E$-field response to the helical pitch modulation is remarkably smaller (up to 0.14 V μm$^{-1}$) than that of the reported system (Figure 5n).[31,32,24]

## 3. Conclusion

In conclusion, we developed just "straight" polar rod mesogens, **nBOE** (n = 1–8), in which the dipole moment aligned nearly perfectly parallel to the molecular axis. Unlike the characteristics of the $N_F$ phase, which emerges in a library of over 150 types of molecules, the enantiotropic $N_F$ phase was observed even in **nBOE** molecules with long alkyl chains (up to n =6). For **nBOE** (n = 4–6) with medium-length alkyl chains, we discovered emergent $^{HC}N_F$ and $SmC_F$ phases with heliconical structures and small molecular tile angles, respectively. The DR, $P–E$, and SHG studies evidenced the ferroelectricity of the $N_F$, $^{HC}N_F$, and $SmC_F$ phases owing to its giant dielectric permittivity (6k–8k), large spontaneous polarization (4.6–6.5 μC cm$^{-2}$), and high SHG activation. The findings from XRD and spectra analysis elucidated that the exotic phase sequence ($N_F \rightarrow {}^{HC}N_F \rightarrow SmC_F$) proceeds via a mechanism that eliminates the offset level between adjacent molecules due to alterations in the strength of multipoint interactions. Additionally, we demonstrated ultralow $E$-field-driven color tunability across the entire VIS-NIR spectral range of the $^{HC}N_F$ phase. We believe that the straight polar rod model can be utilized as a novel strategy for the emergence of spontaneous polar and chiral symmetry breaking**,** unlocking novel helielectric phases in polar fluid materials.

## Supporting Information

Supporting Information is available from the Wiley Online Library or from the author.

## Acknowledgements

We are grateful to Dr. Y. Ishida (RIKEN, CEMS), Dr. H. Koshino (RIKEN, CSRS), and Dr. H. Sato (RIKEN, CEMS), Prof. H. Kikuchi (Kyushu University, IMCE) for allowing us to use a NANOPIX 3.5m system (Rigaku), JNM-ECZ500 (500 MHz, JEOL), and QTOF compact (BRUKER), DSC1 (Mettler Toledo), respectively. We would like to acknowledge the Hokusai GreatWave Supercomputing Facility (project no. RB230008) at the RIKEN Advanced Center for assistance in computing and communication. This work was partially supported by JSPS KAKENHI (JP22K14594; H.N., JP21K14605, JP23H01942; D.O.; JP21H01801, JP23K17341; F.A.), RIKEN Special Postdoctoral Researchers (SPDR) fellowship (H.N.), RIKEN Incentive

**Conflict of Interest**

The authors declare no conflict of interest.

**Author Contributions**

H.N. conceived the project and designed the experiments. F.A. co-designed the work and constructed the optical and electrical setups for SHG. H.N. performed all the experiments. D.O constructed the optical setups and performed the optical experiments. F.A supported XRD measurements. D.K. partially performed $P$–$E$ hysteresis/DR studies. H.N. and D.K. partially synthesized compounds. A.N. synthesized all compounds. M.K. co-designed a synthetic strategy and measured HRMS. M.H. measured and analyzed single crystal XRD. H.N. and F.A. analyzed data and discussed the results. H.N. and F.A wrote the manuscript, and all authors approved the final manuscript.

**Data Availability Statement**

The authors declare that the data supporting the findings of this study are available within the paper and its supplementary information files. All other information is available from the corresponding authors upon reasonable request.

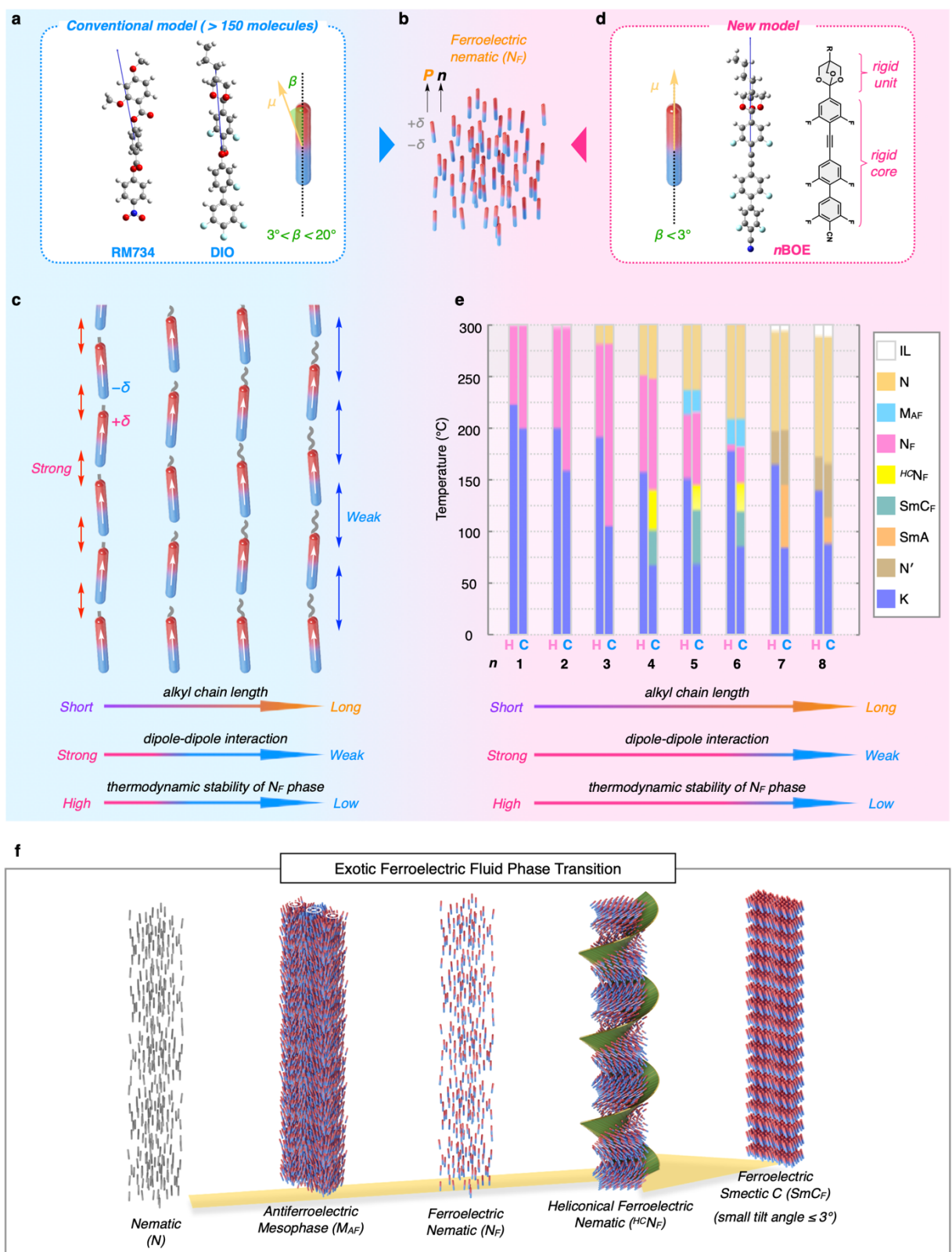

**Figure 1.** A schematic illustration of a ferroelectric nematic ($N_F$) phase (b) induced by conventional (RM734, DIO) (a) and straight rod mesogen (**BOE**) (d). $\beta$ denotes the angle between dipole moment and a direction of long axis for the molecule. c) A plausible model to induce the $N_F$ phase. e) Phase transition behavior of **nBOE** (n = 1–8). Gray-colored area represents decomposition temperature which investigated by thermogravimetric analysis.

Abbrev.: IL = isotropic liquid, N = nematic, $M_{AF}$ = antiferroelectric mesophase, $N_F$ = ferroelectric nematic, $^{HC}N_F$ = heliconical ferroelectric nematic, $SmC_F$ = ferroelectric smectic C, SmA = smectic A, N′ = unknown nematic (apolar), K = crystal. f) A schematic illustration of an exotic ferroelectric phase transition of **BOE** system. The unique interaction in **BOE** lead to not only $N_F$ phase but also heliconical ferroelectric nematic ($^{HC}N_F$) and ferroelectric smectic C ($SmC_F$) phases.

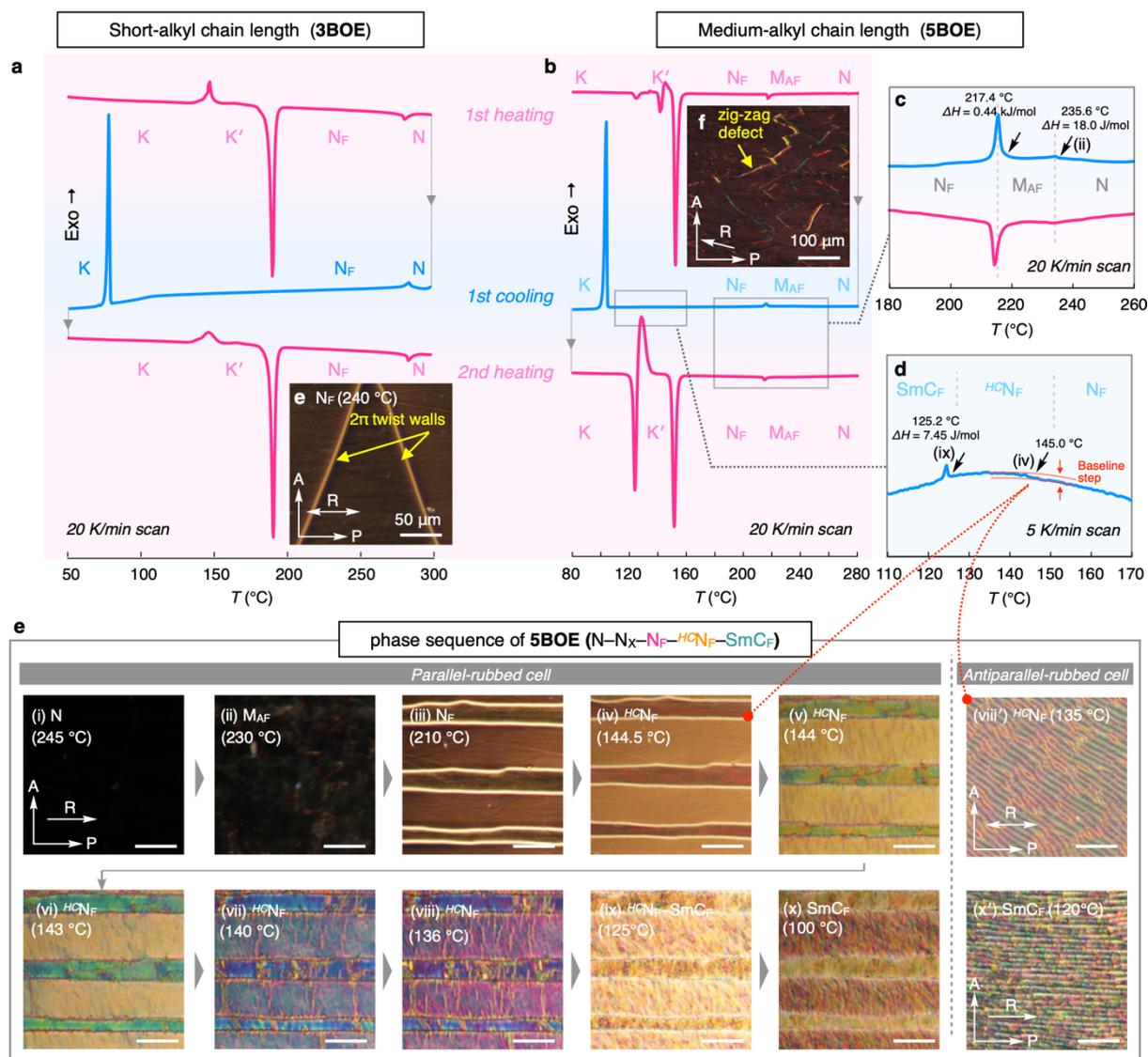

**Figure 2.** DSC studies. DSC curves of **3BOE** (a) and **5BOE** (b–d). Scan rate: 20 K min⁻¹ (a,b,c), 5 K min⁻¹ (d). e) POM images taken under cross polarizers in the parallel- and antiparallel rubbed cells. bar: 50 μm. Areas designated by Roman numerals (i–x, viii′ and x′) corresponds to the observations in DSC and POM studies.

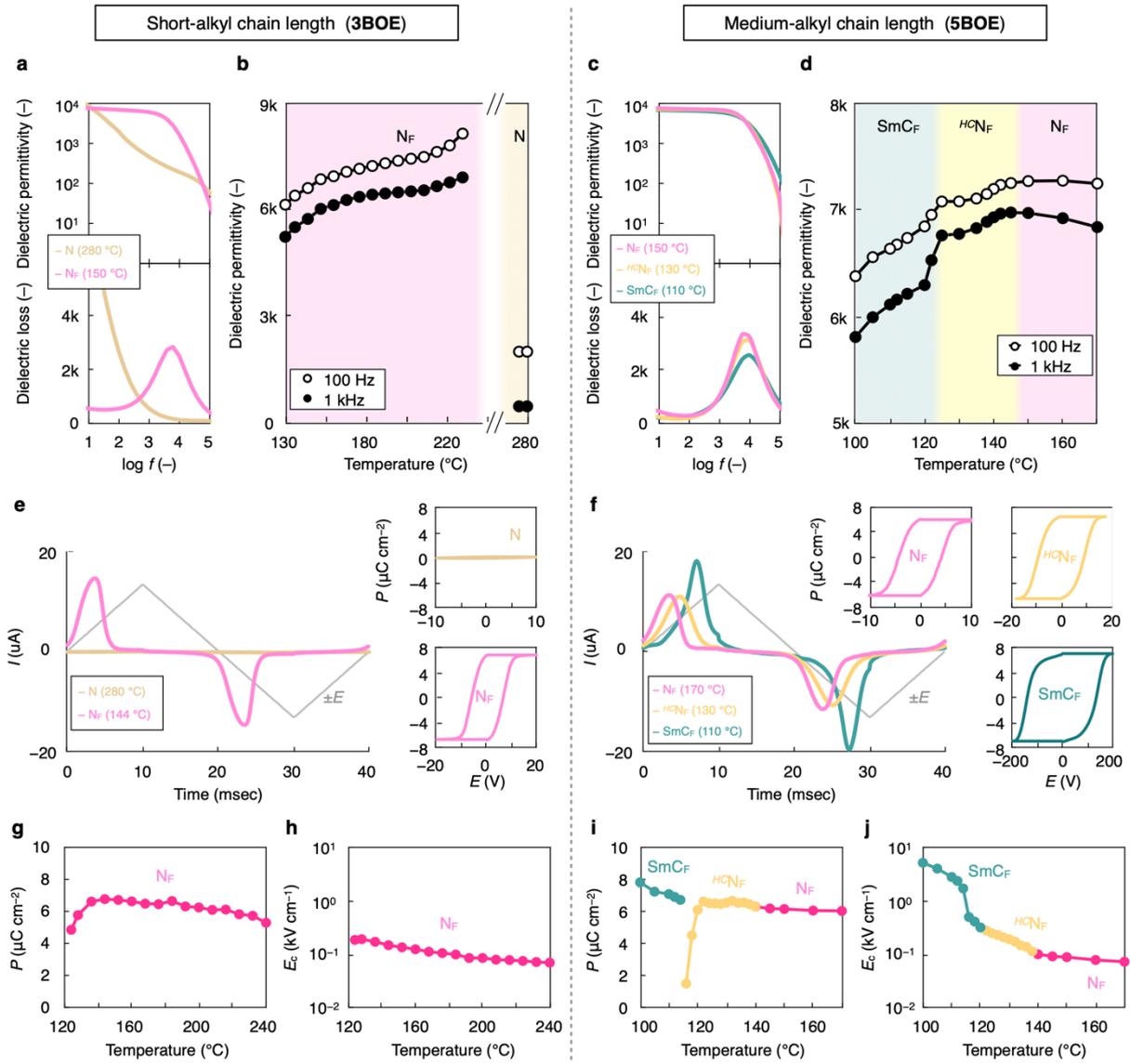

**Figure 3** Polarization behavior. Dielectric spectra (**3BOE**: a, **5BOE**: c) and dielectric permittivity vs temperature (**3BOE**: b, **5BOE**: d). Polarization reversal current profile (**3BOE**: e, **5BOE**: f). Insets represent the corresponding $P$–$E$ hysteresis curves. Polarization density vs temperature (**3BOE**: g, **5BOE**: i). Coercive electric field ($E_c$) vs temperature (**3BOE**: h, **5BOE**: j).

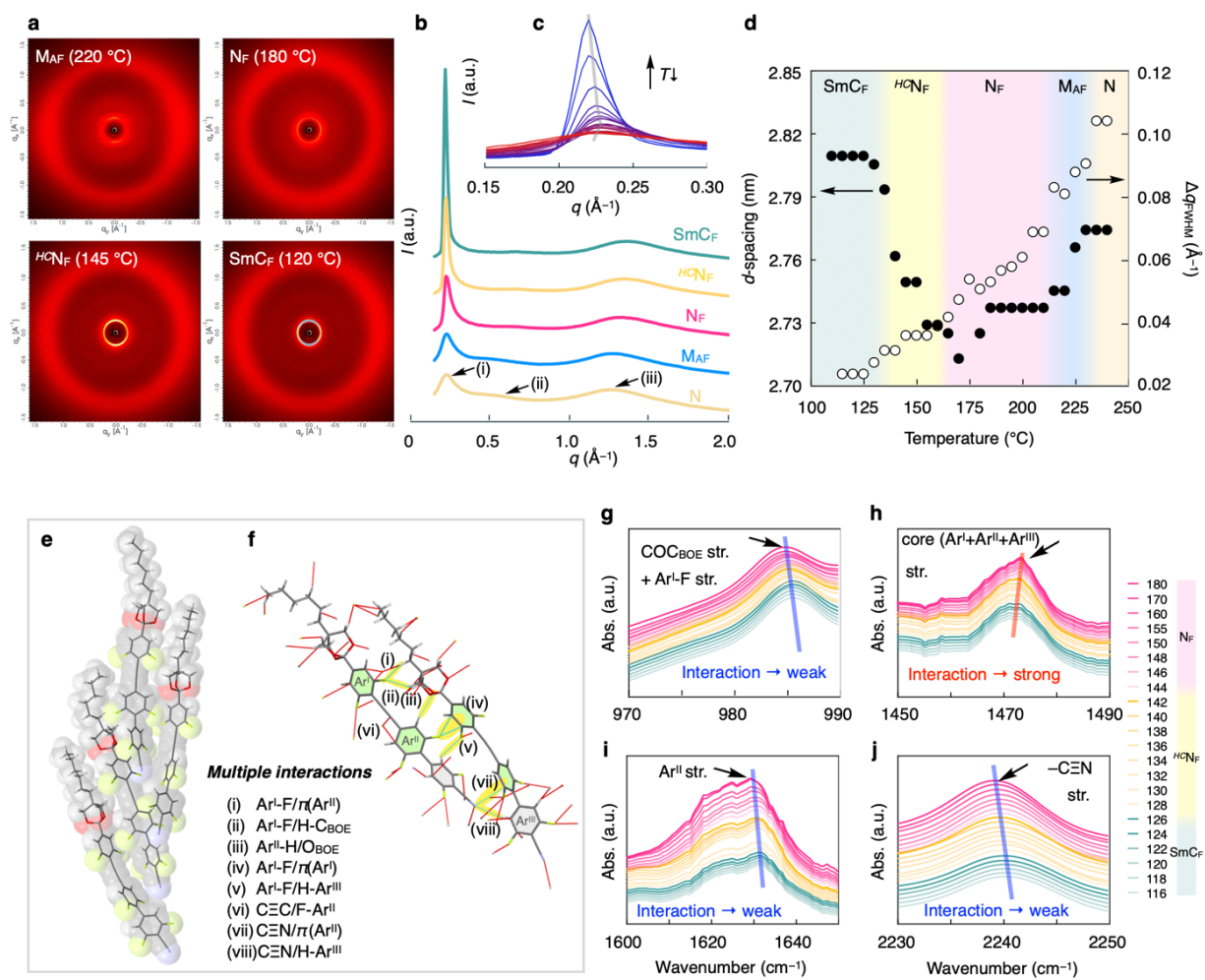

**Figure 4.** XRD and FTIR spectra studies. 2D XRD (a) and 1D XRD (b,c) pattern in various phases for **6BOE**. (d) *d*-spacing and FWHM vs Temperature for **6BOE**. Structures of local cluster (e) and adjacent dimer (f) extracted from SC-XRD data for **6BOE**. (g–j) FTIR absorbance spectra for **5BOE**. For the panel (b), the recorded temperature in N, $N_F$, $^{HC}N_F$ and $SmC_F$ phases are 240, 220, 180, 145 and 120 °C, respectively.

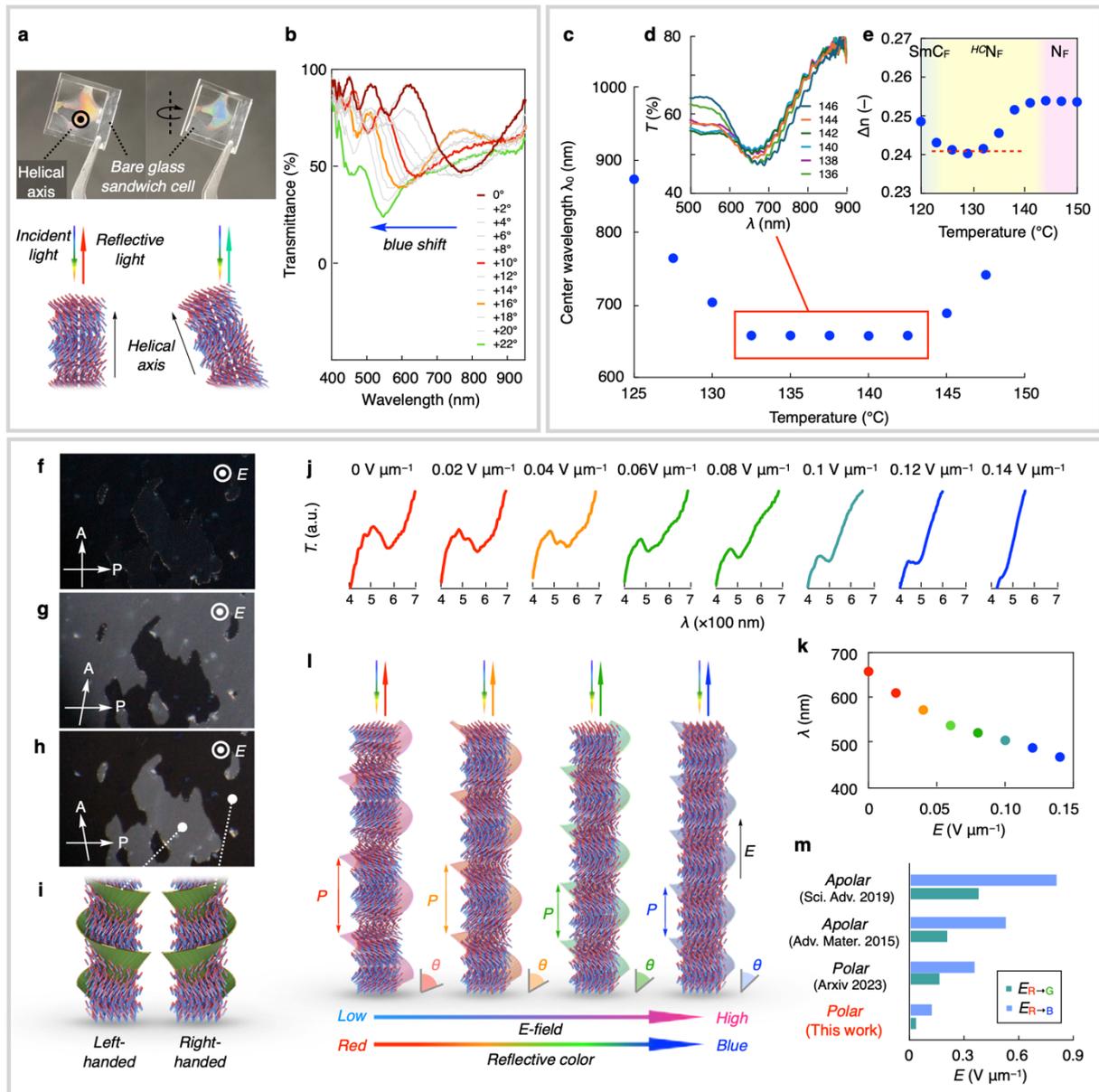

**Figure 5** Unique properties of $^{HC}N_F$ phase emerging in **5BOE**. a) Rotation angle dependence of reflection spectra in the $^{HC}N_F$ phase (145 °C). b) Real photos of $^{HC}N_F$ phase (~145 °C) By rotating the cell, the reflection color changed from orange to blue. c) A schematic illustration of helicoidal structure of $^{HC}N_F$ and estimated color of reflective light. d) Center wavelength ($\lambda$) of reflection band vs Temperature. e) Birefringence vs temperature, POM images during an applied *E*-field (0.25 V μm$^{-1}$) under crossed (f) and decrossed polarizers (g,h). i) left-/right-handed heliconical ferroelectric nematic structure. *E*-field driven color modulation of a $^{HC}N_F$LC film (10 μm) (j) and the corresponding peak shift (k). l) Schematic illustration of ferroelectric heliconical structure and their structure change under *E*-field. m) Comparison of required *E*-field to change in reflective color from red to green or to blue. In the panel (m), examples of conventional apolar and latest polar heliconical system are shown.

# Emergent Ferroelectric Nematic and Heliconical Ferroelectric Nematic States in an Achiral "Straight" Polar Rod Mesogen

Hiroya Nishikawa*, Daichi Okada, Dennis Kwaria, Atsuko Nihonyanagi, Motonobu Kuwayama, Manabu Hoshino and Fumito Araoka*

*To whom correspondence should be addressed.

E-mail: hiroya.nishikawa@riken.jp (H.N.) and fumito.araoka@riken.jp (F.A.)

**Table of contents**

**Methods**

**Methods**

**1. General and materials**

**Nuclear magnetic resonance (NMR) spectroscopy**: $^1$H, $^{13}$C, and $^{19}$F NMR spectra were recorded on JNM-ECZ500 (JEOL) operating at 500 MHz, 126 MHz, and 471 MHz for $^1$H [$^1$H{$^{19}$F}], $^{13}$C{$^1$H} [$^{13}$C{$^1$H,$^{19}$F}] and $^{19}$F [$^{19}$F{$^1$H}] NMR, respectively, using the TMS (trimethylsilane) as an internal standard for $^1$H NMR and the deuterated solvent for $^{13}$C NMR. The absolute values of the coupling constants are given in Hz, regardless of their signs. Signal multiplicities were abbreviated by s (singlet), d (doublet), t (triplet), q (quartet), quint (quintet), sext (sextet), and dd (double–doublet), respectively.

**High-resolution mass (HRMS) spectroscopy**: The quadrupole time-of-flight high-resolution mass spectrometry (QTOF-HRMS) was performed on COMPACT (BRUKER). The calibration was carried out using LC/MS tuning mix, for APCI/APPI (Agilent Technologies).

**Density Functional Theory (DFT) Calculation**: Calculations were performed using the Chem3D (pro, 22.2.0.3300) and Gaussian 16 (G16, C.01) softwares (installed at the RIKEN Hokusai GreatWave Supercomputing facility) for MM2 and DFT calculations, respectively. GaussView 6 (6.0.16) software was used to visually analyze the calculation results. Positions of hydrogens of molecules were optimized using the B3LYP/6-31G++ level Gaussian 16 program package. [S1] Dipole moments of molecules were calculated using b3lyp/6-311+g(d,p) and B3LYP-gd3bj which were added for empirical dispersion corrections to the standard B3LYP. The calculation method is as follows: opt=tight freq b3lyp/6-311+g(d,p) geom=connectivity empiricaldispersion=gd3bj int=ultrafine.

**Polarized optical microscopy**: Polarized optical microscopy was performed on a polarizing microscope (Eclipse LV100 POL, Nikon) with a hot stage (HSC402, INSTEC) on the rotation stage. Unless otherwise noted, the sample temperature was controlled using the INSTEC temperature controller and a liquid nitrogen cooling system pump (mk2000 and LN2-P/LN2-D2, INSTEC).

**Differential scanning calorimetry (DSC).** Differential scanning calorimetry was performed on a calorimeter (DSC1, Mettler-Toledo). Rate, 5 and 20 K min$^{-1}$. Cooling/heating profiles were recorded and analyzed using the Mettler-Toledo STARe software system.

**Dielectric spectroscopy.** Dielectric relaxation spectroscopy was performed ranging between 1 Hz and 1 MHz using an impedance/gain-phase analyzer (SI 1260, Solartron Metrology) and a dielectric interface (SI 1296, Solartron Metrology). Prior to starting the measurement of the LC sample, the capacitance of the empty cell was determined as a reference.

**P–E hysteresis measurement.** *P–E* hysteresis measurements were performed in the temperature range of the $N_F$ phase under a triangular-wave electric field (10 kV cm$^{-1}$, 200 Hz) using a ferroelectricity evaluation system (FCE 10, TOYO Corporation), which is composed of an arbitrary waveform generator (2411B), an IV/QV amplifier (model 6252) and a simultaneous A/D USB device (DT9832).

**SHG measurement.** The SHG investigation was carried out using a Q-switched DPSS Nd: YAG laser (FQS-400-1-Y-1064, Elforlight) at $\lambda$ = 1064 nm with a 5 ns pulse width (pulse energy: 400 μJ). The primary beam was incident on the LC cell followed by the detection of the SHG signal. The electric field was applied normally to the LC cell. The optical setup is shown in Figure S1. The experimental details are mentioned in Supplementary Note 1.

**Wide-angle X-ray scattering (WAXD) measruement.** Two-dimensional WAXD measurement was carried out using the NANOPIX system (Rigaku). The samples held in a Cu holder (diameter: 2.0 mm, thickness: 2 mm) were measured at a constant temperature using a temperature controller (mk2000B, INSTEC) and a hot stage. The measurement set up is shown in Figure S2. The scattering vector $q$ ($q = 4\pi\sin\theta\,\lambda^{-1}$; $2\theta$ and $\lambda$ = scattering angle and wavelength of an incident X-ray beam [1.54 Å (NANOPIX)] and position of an incident X-ray beam on the detector were calibrated using several orders of layer diffractions from silver behenate ($d$ = 58.380 Å). The sample-to-detector distances were 95.4 mm (NANOPIX), where acquired scattering 2D images were integrated along the Debye–Scherrer ring by using software (Igor Pro with Nika-plugin), affording the corresponding one-dimensional profiles.

**Single crystal X-ray scattering (SC-XRD) measurement.**

A single crystal of **6BOE** with dimensions of 0.29 × 0.09 × 0.04 mm was used for SC-XRD measurement. Diffraction data were collected using Rigaku Synergy-i single crystal X-ray diffractometer equipped with a sealed-tube CuK$\alpha$ ($\lambda$ = 1.54184 Å) X-ray source and a pixel array type detector. The single crystal was mounted on the goniometer head of the diffractometer and cooled to 100 K by a cold nitrogen stream from Oxford Cryosystem Cryostream 800 for measurement. Collected data were integrated, corrected (including the Lorentzian-polarization and the absorption correction), and scaled using the program *CrysAlisPro*.[S2] All crystal structures were solved using the program *SHELXT*[S3] and refined (against $|F^2|$)using the program *SHELXL*.[S4] Crystal system: monoclinic. Space group $P2_1/c$. Unit cell parameters: $a = 29.5406(2)$ Å, $b = 25.3902(2)$ Å, $c = 14.49345(13)$ Å, $\beta = 91.1515(7)°$, $V = 10868.49(15)$ Å$^3$. Calculated Density: 1.431 g cm$^{-1}$. $2\theta_{max}$: 134.16°. Scan mode: $\omega$ scans. No. of reflections: 72370 (measured) and 28833 (independent). $\mu$: 1.026 mm$^{-1}$. $T_{min}$, $T_{max}$: 0.510, 1.000. No. of parameters: 1518. $Z = 16$. $R$ ($F^2 > 2\sigma(F^2)$) = 0.0771, $wR$ (all reflections) = 0.1808. and Residual electron density: 0.872(max), $-0.257$(min). CCDC reference number: 2356894.

**Reflection Spectra measurement.** The reflection spectra were recorded using a fiber optic spectrometer (USB4000, Ocean optics). The rotation angle dependence of reflection spectra was recorded using the optical setup as shown in Figure S3. For experiments on reflection color tuning, we employed a multifunction generator (WF1973, nf) and a microscope spectroscopic method using the polarizing microscope (Eclipse LV100 POL, Nikon) quipped with the fiber optic spectrometer (USB4000, Ocean optics).

**Birefringence measurement.** Birefringence ($\Delta$n) measurement was performed using the microscope spectroscopic method using the polarizing microscope (Eclipse LV100 POL, Nikon) quipped with the fiber optic spectrometer (USB4000, Ocean optics). The transmittance of light under para-polarization conditions was observed. The obtained transmittance data was fitted by combination of eq. (1) and Cauchy's eq. (2),

$$I = I_0 cos^2\left(\frac{\pi d\Delta n(\lambda)}{\lambda}\right) \qquad (1)$$

$$\Delta n(\lambda) = A + B/\lambda^2 + C/\lambda^4 \qquad (2)$$

, where $I$ and $\lambda$ are intensity of transmitted light, wavelength, respectively while, $A$, $B$ and $C$ are Cauchy's coefficients. In this paper, we adopted $\Delta$n(550).

**FTIR spectra measurement.** FTIR absorbance spectra were recorded using AIM-9000 FTIR microscope systems (SHIMADZU) with simultaneous observation of light microscope images. The LC sample was injected in a BaF$_2$ cell. The corresponding FTIR spectra was simulated by DFT calculation as mentioned above.

**Information of used liquid crystalline (LC) cells**:

*Bare glass cell (EHC)*:

  - Experiments: POM (thickness: 10.0 μm) and spectra studies (thickness: 5.0, 10.0 μm)

*Bare glass cell (homemade)*:

  - Sandwich-type film using two cover glasses

  - Experiments: Spectra studies (thickness: 10.0 μm)

*Antiparallel-rubbed cell (EHC)*:

  - PI-coated type

  - Alignment layer: LX-1400

  - Experiments: POM studies (thickness: 2.0, 5.0 μm)

*Parallel-rubbed cell (EHC)*:

  - PI-coated type

  - Alignment layer: LX-1400

  - Experiments: POM studies (thickness: 2.0, 5.0 μm), birefringence measurement (thickness: 2.0 μm).

*ITO glass cell (EHC)*:

  - ITO-coated type, electrode area: 5 × 10 mm

  - Experiments: POM (thickness: 10.0 μm) and DR (thickness: 9.0 μm) studies; reflection color tuning (thickness: 10.0 μm).

*IPS cell (EHC)*:

  - PI-coated type, electrode distance: 500 μm, electrode length: 18 mm

  - Alignment layer: LX-1400

  - Rubbing condition: antiparallel

  - Experiments: *PE* hysteresis (thickness: 5.0 μm)

*IPS cell (homemade)*:

  - Au-coated type (Cr bottom layer: 5 nm; Au top layer: 50 nm), electrode distance: 1 mm, electrode length: 13 mm, electrode width, 2 mm

  - Alignment layer: AL1254

  - Rubbing condition: antiparallel (rubbing depth: 100 μm, rubbing times: 10)

  - Experiments: SHG (thickness: 5.0 μm) studies

## 2. Synthesis of nBOE (n = 1–8).

## 2.1. Synthetic route

**nBOE** (n = 1–8) used in this paper were synthesized by following pathway (**Scheme S1**). The blue and green colored pathway indicate solution chemical (SC) and mechanochemical (MC) synthesis, respectively. The detail of MC synthesis has been reported in our previous papers.[S5,S6]

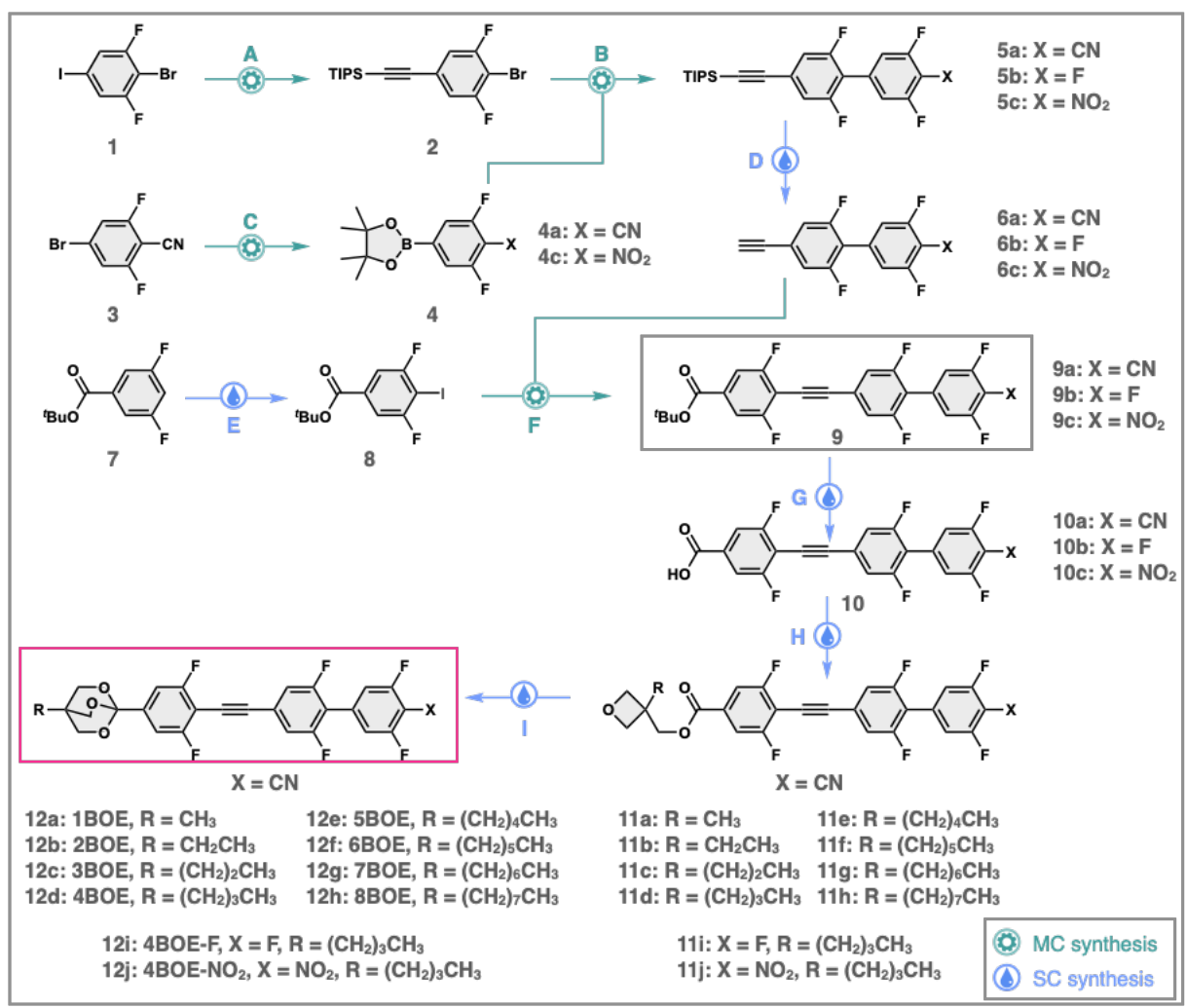

**Scheme S1** Synthetic pathway of **nBOE** (n = 1–8), **4BOE-F** and **4BOE-NO₂**. A) MC-Sonogashira coupling, B) MC-Suzuki coupling, C) MC-Miyaura coupling, D) deprotection of TIPS group with TBAF, E) Iodization, F) MC-Sonogashira coupling, G) deprotection of ᵗBuO group, H) esterification, I) orthoesterification.

## 2.2. SC/MC Synthesis

### 2.2.1. *Synthesis of ((4-bromo-3,5-difluorophenyl)ethynyl)triisopropylsilane (2)*

Compound **2** was synthesized by following the procedure described in our previous paper.[S5]

### 2.2.2. *Synthesis of compounds (4a–c)*

Compounds **4a–c** were synthesized by following the procedure described in our previous paper.[S5]

#### *2,6-difluoro-4-(4,4,5,5-tetramethyl-1,3,2-dioxaborolan-2-yl)benzonitrile (4a)*

The $^1$H NMR spectrum is in accordance with the literature.[S5]

#### *2-(3,5-Difluoro-4-nitrophenyl)-4,4,5,5-tetramethyl-1,3,2-dioxaborolane (4c)*

$^1$H-NMR (500 MHz, CDCl$_3$): δ 7.48 (d, J = 8.0 Hz, 2H), 1.35 (s, 12H)

### 2.2.3. *Synthesis of compounds (5a–c)*

Compounds **5a–c** were synthesized by following the procedure described in our previous paper.[S5]

#### *2′,3,5,6′-Tetrafluoro-4′-((triisopropylsilyl)ethynyl)-[1,1′-biphenyl]-4-carbonitrile (5a)*

The $^1$H NMR spectrum is in accordance with the literature.[S5]

#### *Triisopropyl((2,3′,4′,5′,6-pentafluoro-[1,1′-biphenyl]-4-yl)ethynyl)silane (5b)*

$^1$H-NMR (500 MHz, CDCl$_3$): δ 7.12-7.09 (m, 4H), 1.13 (s, 18H)

#### *Triisopropyl((2,3′,5′,6-tetrafluoro-4′-nitro-[1,1′-biphenyl]-4-yl)ethynyl)silane (5c)*

$^1$H-NMR (500 MHz, CDCl$_3$): δ 7.24 (d, J = 8.5 Hz, 2H), 7.14 (d, J = 8.5 Hz, 2H), 1.14 (s, 18H)

### 2.2.4. *General procedure of compounds (6a–c)*

Compounds **6a–c** were synthesized by following the procedure described in our previous paper.[S5]

#### *4′-Ethynyl-2′,3,5,6′-tetrafluoro-[1,1′-biphenyl]-4-carbonitrile (6a)*

The $^1$H NMR spectrum is in accordance with the literature.[S5]

#### *4-Ethynyl-2,3′,4′,5′,6-pentafluoro-1,1′-biphenyl (6b)*

$^1$H-NMR (500 MHz, CDCl$_3$): δ 7.15-7.10 (m, 4H), 3.22 (s, 1H)

#### *4-Ethynyl-2,3′,5′,6-tetrafluoro-4′-nitro-1,1′-biphenyl (6c)*

$^1$H-NMR (500 MHz, CDCl$_3$): δ 7.25 (d, J = 8.5 Hz, 2H), 7.19-7.15 (m, 2H), 3.28 (s, 1H)

### 2.2.5. *Synthesis of tert-butyl 3,5-difluoro-4-iodobenzoate (8)*

LDA was prepared by adding n-BuLi (27.6 mL, 43 mmol, 1.56 M in hexane) to a solution of diisopropylamine (6.0 mL, 43 mmol) in 200 mL of THF at –78 °C under Ar. The solution was stirred 30 min at 0 °C and then cooled again to –78 °C. To the LDA solution was added a solution of tert-butyl 3,5-difluorobenzoate (7.7 g, 35.8 mmol) in THF (10 mL), and the mixture was stirred for 30 min. Iodine (10 g, 39.4 mmol) was then added to the mixture. After 5 min stirring, the reaction was warmed to 0 °C and stirred for 1.5 h. The mixture was poured into ice water and extracted with EtOAc and hexane (1:1), the organic layer was dried over $Na_2SO_4$. The solution was filtered through a short pad of silica gel and the short pad was further washed with mixed solvent (hexane / $CH_2Cl_2$ =1:1). The solution was concentrated and dried under reduced pressure to yield a brown solid (9.8 g, 28.7 mmol, 80.2%)

[1]H-NMR (600 MHz, CDCl$_3$): δ 7.47 (2H, d, $J$ = 6.6 Hz), 1.58 (9H, s)

### *Synthesis of compounds (9a–c)*

Compounds **9a–c** were synthesized by following the procedure described in our previous paper.[S5].

#### *tert-Butyl 4-((4'-cyano-2,3',5',6-tetrafluoro-[1,1'-biphenyl]-4-yl)ethynyl)-3,5-difluorobenzoate (9a)*

[1]H-NMR (500 MHz, CDCl$_3$): δ 7.57 (d, J = 7.5 Hz, 2H), 7.28 (d, J = 5.5 Hz, 2H), 7.23 (d, J = 8.0 Hz, 2H), 1.60 (s, 9H)

#### *tert-Butyl 3,5-difluoro-4-((2,3',4',5',6-pentafluoro-[1,1'-biphenyl]-4-yl)ethynyl)benzoate (9b)*

[1]H-NMR (500 MHz, CDCl$_3$): δ 7.57 (d, J = 7.5 Hz, 2H), 7.24 (d, J = 7.5 Hz, 2H), 7.15 (t, J = 7.3 Hz, 2H), 1.60 (s, 9H)

#### *tert-Butyl 3,5-difluoro-4-((2,3',5',6-tetrafluoro-4'-nitro-[1,1'-biphenyl]-4-yl)ethynyl)benzoate (9c)*

[1]H-NMR (500 MHz, CDCl$_3$): δ 7.58 (d, J = 7.5 Hz, 2H), 7.30-7.26 (m, 4H), 1.61 (s, 9H)

### 2.2.6. Synthesis of compounds (10a–c)

*Example:* To a solution of *tert*-butyl ester (1.57 g, 3.22 mmol) in $CH_2Cl_2$ (18 mL) was added TFA (6 mL) at 0 °C under Ar. After stirring for 1h, the resulting solution was warmed to r.t. and further stirred for 3.5 h. The reaction was quenched by the addition of $H_2O$ and extracted with $CH_2Cl_2$. The combined organic layer was dried over $Na_2SO_4$ and concentrated in vacuo afforded **10a** as a pale brown solid (1.34 g, 3.11 mmol, 96.6 %). The other compounds were synthesized according to above procedure.

### 4-((4'-Cyano-2,3',5',6-tetrafluoro-[1,1'-biphenyl]-4-yl)ethynyl)-3,5-difluorobenzoic acid (10a)

$^1$H{$^{19}$F}-NMR (500 MHz, acetone-$d6$): δ 7.73 (s, 2H), 7.61 (s, 2H), 7.53 (s, 2H)

### 3,5-Difluoro-4-((2,3',4',5',6-pentafluoro-[1,1'-biphenyl]-4-yl)ethynyl)benzoic acid (10b)

$^1$H-NMR (500 MHz, CD$_3$OD): δ 7.68 (d, J = 7.6 Hz, 2H), 7.35 (m, 4H)

### 3,5-Difluoro-4-((2,3',5',6-tetrafluoro-4'-nitro-[1,1'-biphenyl]-4-yl)ethynyl)benzoic acid (10c)

$^1$H-NMR (500 MHz, acetone-$d6$): δ 7.73 (d, J = 8.0 Hz, 2H), 7.68 (d, J = 9.0 Hz, 2H), 7.53 (d, J = 8.0 Hz, 2H)

## 2.2.7. Synthesis of compounds (11a–c)

*Example:* To a solution of carboxylic acid (620 mg, 1.44 mmol) in CH$_2$Cl$_2$ (17 mL) were added 3-pentyl-3-oxetanemethanol (266 mg, 1.68 mmol), EDAC-HCl (322 mg, 1.68 mmol), and DMAP (51.3 mg, 0.42 mmol) at 0 °C under Ar. After 0.5 h stirring, the solution was warmed to r.t. and further stirred for 2 h. The reaction was quenched by the addition of H$_2$O and extracted with CH$_2$Cl$_2$. The combined organic layer was dried over Na$_2$SO$_4$ and concentrated. The crude products were purified by column chromatography on silica gel (hexane/ CH$_2$Cl$_2$, 1:4 to 0:100) afforded **11e** as a white solid (360 mg, 0.64 mmol, 99%). The other compounds were synthesized according to above procedure.

### (3-Methyloxetan-3-yl)methyl 3,5-difluoro-4-((2,3',5',6-tetrafluoro-4'-nitro-[1,1'-biphenyl]-4-yl)ethynyl)benzoate (11a)

$^1$H{$^{19}$F}-NMR (500 MHz, CDCl$_3$): δ 7.65 (s, 2H), 7.29 (s, 2H), 7.23 (s, 2H), 4.56 (dd, J = 62.6, 6.0 Hz, 4H), 4.45 (s, 2H), 1.44 (s, 3H)

### (3-Ethyloxetan-3-yl)methyl 3,5-difluoro-4-((2,3',5',6-tetrafluoro-4'-nitro-[1,1'-biphenyl]-4-yl)ethynyl)benzoate (11b)

$^1$H{$^{19}$F}-NMR (500 MHz, CDCl$_3$): δ 7.64 (s, 2H), 7.29 (s, 2H), 7.23 (s, 2H), 4.54 (dd, J = 28.8, 6.2 Hz, 4H), 4.50 (s, 2H), 1.85 (q, J = 7.4 Hz, 2H), 0.98 (t, J = 7.5 Hz, 3H)

### (3-Propyloxetan-3-yl)methyl 4-((4'-cyano-2,3',5',6-tetrafluoro-[1,1'-biphenyl]-4-yl)ethynyl)-3,5-difluorobenzoate (11c)

$^1$H{$^{19}$F}-NMR (500 MHz, CDCl$_3$): δ 7.64 (s, 2H), 7.29 (s, 2H), 7.23 (s, 2H), 4.55 (dd, J = 21.8, 6.0 Hz, 4H), 4.50 (s, 2H), 1.80-1.77 (m, 2H), 1.41-1.33 (m, 2H), 0.99 (t, J = 7.5 Hz, 3H)

***(3-Butyloxetan-3-yl)methyl 4-((4'-cyano-2,3',5',6-tetrafluoro-[1,1'-biphenyl]-4-yl)ethynyl)-3,5-difluorobenzoate (11d)***

[1]H-NMR (600 MHz, CDCl$_3$): δ 7.64 (d, J = 7.2 Hz, 2H), 7.29 (d, J = 7.8 Hz, 1H), 7.23 (d, J = 9.0 Hz, 2H), 4.55 (dd, J = 31.8, 6.0 Hz, 4H), 4.49 (s, 2H), 1.82-1.79 (m, 2H), 1.38-1.30 (m, 4H), 0.94 (t, J = 6.9 Hz, 3H)

***(3-Pentyloxetan-3-yl)methyl 4-((4'-cyano-2,3',5',6-tetrafluoro-[1,1'-biphenyl]-4-yl)ethynyl)-3,5-difluorobenzoate (11e)***

[1]H{[19]F}-NMR (500 MHz, CDCl$_3$): δ 7.64 (s, 2H), 7.29 (s, 2H), 7.23 (s, 2H), 4.54 (dd, J = 27.0, 6.0 Hz, 4H), 4.49 (s, 2H), 1.79 (m, 2H), 1.37-1.25 (m, 6H), 0.91 (t, J = 6.8 Hz, 3H)

***(3-Hexyloxetan-3-yl)methyl 4-((4'-cyano-2,3',5',6-tetrafluoro-[1,1'-biphenyl]-4-yl)ethynyl)-3,5-difluorobenzoate (11f)***

[1]H-NMR (500 MHz, CDCl$_3$): δ 7.64 (d, J = 7.0 Hz, 2H), 7.29 (d, J = 8.0 Hz, 1H), 7.23 (d, J = 8.0 Hz, 2H), 4.54 (dd, J = 26.8, 6.2 Hz, 3H), 4.49 (s, 2H), 1.81-1.78 (m, 2H), 1.34-1.30 (m, 8H), 0.89 (t, J = 6.8 Hz, 3H)

***(3-Heptyloxetan-3-yl)methyl 4-((4'-cyano-2,3',5',6-tetrafluoro-[1,1'-biphenyl]-4-yl)ethynyl)-3,5-difluorobenzoate (11g)***

[1]H-NMR (500 MHz, CDCl$_3$): δ 7.64 (d, J = 7.5 Hz, 2H), 7.29 (d, J = 8.0 Hz, 1H), 7.23 (d, J = 8.0 Hz, 2H), 4.54 (dd, J = 26.5, 6.0 Hz, 3H), 4.49 (s, 2H), 1.81-1.79 (m, 2H), 1.32-1.28 (0, 10H), 0.89 (t, J = 7.0 Hz, 3H)

***(3-Octyloxetan-3-yl)methyl 4-((4'-cyano-2,3',5',6-tetrafluoro-[1,1'-biphenyl]-4-yl)ethynyl)-3,5-difluorobenzoate (11h)***

[1]H{[19]F}-NMR (500 MHz, CDCl$_3$): δ 7.64 (s, 2H), 7.29 (s, 2H), 7.23 (s, 2H), 4.54 (dd, J = 26.5, 6.0 Hz, 3H), 4.49 (s, 2H), 1.81-1.78 (m, J = 7.6 Hz, 2H), 1.33-1.27 (m, 12H), 0.88 (t, J = 6.8 Hz, 3H)

***(3-Butyloxetan-3-yl)methyl 3,5-difluoro-4-((2,3',4',5',6-pentafluoro-[1,1'-biphenyl]-4-yl)ethynyl)benzoate (11i)***

[1]H{[19]F}-NMR (500 MHz, CDCl$_3$): δ 7.64 (s, 2H), 7.25 (s, 2H), 7.15 (d, J = 6.0 Hz, 2H), 4.54 (dd, J = 27.0, 6.0 Hz, 3H), 4.49 (s, 2H), 1.82-1.79 (m, 2H), 1.41-1.28 (m, 4H), 0.94 (t, J = 7.3 Hz, 3H)

***(3-Butyloxetan-3-yl)methyl 3,5-difluoro-4-((2,3',5',6-tetrafluoro-4'-nitro-[1,1'-biphenyl]-4-yl)ethynyl)benzoate (11j)***

$^1$H-NMR (500 MHz, CDCl$_3$): δ 7.65 (d, J = 7.0 Hz, 2H), 7.30-7.26 (m, 4H), 4.54 (dd, J = 26.8, 6.3 Hz, 4H), 4.49 (s, 2H), 1.82-1.79 (m, 2H), 1.41-1.28 (m, 4H), 0.94 (t, J = 7.3 Hz, 3H)

### 2.2.8. *Synthesis of nBOE (12a–12h), 4BOE-F (12i) and 4BOE-NO$_2$ (12j)*

*Example:* To a solution of ester (663 mg, 1.19 mmol) in CH$_2$Cl$_2$ (6 mL) was added BF$_3$-OEt$_2$ (14.7 µL, 0.12 mmol) at 0 °C under Ar. After 1.5 h stirring, the solution was warmed to r.t. and further stirred for 1.5 h. The reaction was quenched by the addition of Et$_3$N (828 µL, 5.94 mmol) and H$_2$O, and extracted with CH$_2$Cl$_2$. The combined organic layer was dried over Na$_2$SO$_4$ and concentrated. The residue was purified by a neutral alumina column chromatography on silica gel (hexane / CHCl$_3$ 1:4 to 1:1) afforded **11d** as a white solid (512 mg, 0.92 mmol, 77%). The other compounds were synthesized according to above procedure. All compounds were recrystallized twice from hexane/CH$_2$Cl$_2$.

### *4'-((2,6-Difluoro-4-(4-methyl-2,6,7-trioxabicyclo[2.2.2]octan-1-yl)phenyl)ethynyl)-2',3,5,6'-tetrafluoro-[1,1'-biphenyl]-4-carbonitrile (12a)*

The $^1$H NMR spectrum is in accordance with the literature.[S5]

### *4'-((2,6-Difluoro-4-(4-ethyl-2,6,7-trioxabicyclo[2.2.2]octan-1-yl)phenyl)ethynyl)-2',3,5,6'-tetrafluoro-[1,1'-biphenyl]-4-carbonitrile (12b)*

The $^1$H NMR spectrum is in accordance with the literature.[S5]

### *4'-((2,6-Difluoro-4-(4-propyl-2,6,7-trioxabicyclo[2.2.2]octan-1-yl)phenyl)ethynyl)-2',3,5,6'-tetrafluoro-[1,1'-biphenyl]-4-carbonitrile, 3BOE (12c)*

$^1$H{$^{19}$F}-NMR (500 MHz, CDCl$_3$): 7.25 (s, 2H), 7.24 (s, 2H), 7.23 (s, 2H), 4.11 (s, 6H), 1.31-1.22 (m, 4H), 0.94 (t, J = 6.7 Hz, 3H)

$^{13}$C{$^1$H}-NMR (126 MHz, CDCl$_3$): 162.8, 162.5, 159.1, 141.2, 136.6, 125.9, 115.6, 115.5, 114.2, 109.4, 108.9, 106.3, 101.6, 96.0, 92.3, 80.0, 72.1, 33.5, 31.9, 16.6, 14.7

QTOF-HRMS (*m/z*, [M+H]$^+$) calc for 544.1348; found, 544.1370

### *4'-((2,6-Difluoro-4-(4-butyl-2,6,7-trioxabicyclo[2.2.2]octan-1-yl)phenyl)ethynyl)-2',3,5,6'-tetrafluoro-[1,1'-biphenyl]-4-carbonitrile, 4BOE (12d)*

$^1$H{$^{19}$F}-NMR (500 MHz, CDCl$_3$): 7.25 (s, 2H), 7.24 (s, 2H), 7.23 (s, 2H), 4.11 (s, 6H), 1.35-1.19 (m, 6H), 0.92 (t, J = 7.2 Hz, 3H)

$^{13}$C{$^1$H}-NMR (126 MHz, CDCl$_3$): 162.8, 162.5, 159.1, 141.2, 136.6, 125.9, 115.6, 115.5, 114.2, 109.4, 109.0, 106.3, 101.6, 96.0, 92.3, 80.1, 72.1, 33.4, 29.4, 25.3, 23.3, 13.8

QTOF-HRMS (*m/z*, [M+H]$^+$) calc for 558.1504; found, 558.1502.

***4'-((2,6-Difluoro-4-(4-pentyl-2,6,7-trioxabicyclo[2.2.2]octan-1-yl)phenyl)ethynyl)-2',3,5,6'-tetrafluoro-[1,1'-biphenyl]-4-carbonitrile, 5BOE (12e)***

$^1$H{$^{19}$F}-NMR (500 MHz, CDCl$_3$): 7.25 (s, 2H), 7.24 (s, 2H), 7.23 (s, 2H), 4.11 (s, 6H), 1.35-1.24 (m, 8H), 0.90 (t, J = 7.1 Hz, 3H)

$^{13}$C{$^1$H}-NMR (126 MHz, CDCl$_3$): 162.8, 162.5, 159.1, 141.2, 136.6, 125.9, 115.6, 115.5, 114.2, 109.4, 109.0, 106.3, 101.6, 96.0, 92.3, 80.1, 72.1, 33.5, 32.3, 29.6, 22.8, 22.3, 13.9

QTOF-HRMS (*m/z*, [M+H]$^+$) calc for 572.1660; found, 572.1669

***4'-((2,6-Difluoro-4-(4-hexyl-2,6,7-trioxabicyclo[2.2.2]octan-1-yl)phenyl)ethynyl)-2',3,5,6'-tetrafluoro-[1,1'-biphenyl]-4-carbonitrile, 6BOE (12f)***

$^1$H{$^{19}$F}-NMR (500 MHz, CDCl$_3$): 7.25 (s, 2H), 7.24 (s, 2H), 7.23 (s, 2H), 4.10 (s, 6H), 1.32-1.21 (m, 10H), 0.89 (t, J = 6.8 Hz, 3H)

$^{13}$C{$^1$H}-NMR (126 MHz, CDCl$_3$): 162.8, 162.5, 159.1, 141.2, 136.6, 125.9, 115.6, 115.5, 114.2, 109.4, 109.0, 106.3, 101.6, 96.0, 92.3, 80.1, 72.1, 33.5, 31.5, 29.8, 29.7, 23.1, 22.5, 14.0

QTOF-HRMS (*m/z*, [M+H]$^+$) calc for 586.1817; found, 586.1810

***4'-((2,6-Difluoro-4-(4-heptyl-2,6,7-trioxabicyclo[2.2.2]octan-1-yl)phenyl)ethynyl)-2',3,5,6'-tetrafluoro-[1,1'-biphenyl]-4-carbonitrile, 7BOE (12g)***

$^1$H{$^{19}$F}-NMR (500 MHz, CDCl$_3$): 7.25 (s, 2H), 7.24 (s, 2H), 7.23 (s, 2H), 4.10 (s, 6H), 1.32-1.24 (m, 12H), 0.89 (t, J = 6.9 Hz, 3H)

$^{13}$C{$^1$H}-NMR (126 MHz, CDCl$_3$): 162.8, 162.5, 159.1, 141.2, 136.6, 125.9, 115.6, 115.5, 114.2, 109.4, 109.0, 106.3, 101.6, 96.0, 92.3, 80.1, 72.1, 33.5, 31.7, 30.1, 29.7, 29.0, 23.2, 22.6, 14.1

QTOF-HRMS (*m/z*, [M+H]$^+$) calc for 600.1973; found, 600.1970

***4'-((2,6-Difluoro-4-(4-octyl-2,6,7-trioxabicyclo[2.2.2]octan-1-yl)phenyl)ethynyl)-2',3,5,6'-tetrafluoro-[1,1'-biphenyl]-4-carbonitrile, 8BOE (12h)***

$^1$H{$^{19}$F}-NMR (500 MHz, CDCl$_3$): 7.25 (s, 2H), 7.24 (s, 2H), 7.23 (s, 2H), 4.10 (s, 6H), 1.32-1.24 (m, 14H), 0.89 (t, J = 6.9 Hz, 3H)

$^{13}$C{$^1$H}-NMR (126 MHz, CDCl$_3$): 162.8, 162.5, 159.1, 141.2, 136.6, 125.9, 115.6, 115.5, 114.2, 109.4, 108.9, 106.3, 101.6, 96.0, 92.3, 80.1, 72.1, 33.5, 31.8, 30.2, 29.7, 29.3, 29.2, 23.2, 22.6, 14.1

QTOF-HRMS (*m/z*, [M+H]$^+$) calc for 614.2130; found, 614.2141

***4-Butyl-1-(3,5-difluoro-4-((2,3',4',5',6-pentafluoro-[1,1'-biphenyl]-4-yl)ethynyl)phenyl)-2,6,7-trioxabicyclo[2.2.2]octane, 4BOE-F (12i)***

$^1$H{$^{19}$F}-NMR (500 MHz, CDCl$_3$): δ 7.22 (d, J = 9.3 Hz, 4H), 7.14 (d, J = 6.2 Hz, 2H), 4.10 (s, 6H), 1.35-1.17 (m, 6H), 0.92 (t, J = 7.3 Hz, 3H)

$^{13}$C{$^1$H}-NMR (126 MHz, CDCl$_3$): $^{13}$C{$^1$H}-NMR (126 MHz, CDCl$_3$): δ 162.5, 159.3, 151.0, 140.9, 139.8, 124.5, 124.2, 116.6, 115.3, 114.8, 109.4, 106.4, 101.8, 96.4, 79.2, 72.1, 33.4, 29.4, 25.3, 23.3, 13.8

QTOF-HRMS (*m/z*, [M+H]$^+$) calc for 551.1457; found, 551.1455

**4-Butyl-1-(3,5-difluoro-4-((2,3',5',6-tetrafluoro-4'-nitro-[1,1'-biphenyl]-4-yl)ethynyl)phenyl)-2,6,7-trioxabicyclo[2.2.2]octane, 4BOE-F (12j)**

$^1$H{$^{19}$F}-NMR (500 MHz, CDCl$_3$): δ 7.27 (s, 2H), 7.26 (s, 2H), 7.24 (s, 2H), 4.11 (s, 6H), 1.35-1.19 (m, 6H), 0.92 (t, J = 7.2 Hz, 3H)

$^{13}$C{$^1$H}-NMR (126 MHz, CDCl$_3$): δ 162.5, 159.2, 154.3, 141.2, 134.2, 129.0, 125.9, 115.5, 115.3, 115.0, 109.4, 106.3, 101.6, 96.0, 80.0, 77.2, 77.0, 76.7, 72.1, 33.4, 29.4, 25.3, 23.3, 13.8

QTOF-HRMS (*m/z*, [M+H]$^+$) calc for 578.1402; found, 578.1405

**Supporting Notes (Notes S1–S5)**

**Supporting Note 1 | SHG measurements.**

SHG is a powerful tool that provides detailed studies of polarization switching on surfaces and in bulk systems without being affected by ionic migration. Figure S1 shows the optical setup. In this experiment, we used a laser pulse with a repetition frequency synchronized with the AC triangular wave voltage frequency. This method prevents significant damage to the sample due to ion accumulation on the electrode surface when a direct current (DC) $E$-field is used. Figure S4a and b shows the SH intensity as a function of the waveform phase. In this setup, the 90° and 270° phases correspond to the maximum applied $E$-field with positive and negative polarities, respectively. To obtain a temperature-dependent SHG profile for **nBOE**, we first measured the SHG responses in various LC phases. For example, as to **4BOE**, the SHG profiles of $N_F$ (180 °C), $^{HC}N_F$ (125 °C), and $SmC_F$ (115 °C) were measured in advance. For **8BOE**, the SHG profiles of N (200 °C), N′ (125 °C) and SmA (115 °C) were recorded. We then measured the temperature dependence of the SHG intensity by fixing the phase that showed the maximum intensity at each phase. For **6BOE** and **8BOE**, the fixed phases were 150° and 90°, respectively. As shown in Figure S5a–c, strong SH activity was observed in three phases ($N_F$, $^{HC}N_F$, and $SmC_F$), indicating the presence of a polar structure. Similarly, the SHG activity of **3BOE** was detected (Figure S5d). The data for **1BOE** and **2BOE** were reported in our previous paper[S5]). Furthermore, **8BOE** also exhibited SH activity throughout various phases (N, N′, SmA). However, as seen in Figure S5e, the SH intensity increased and decreased with the absolute value of the $E$-field, suggesting that this response implies paraelectric behavior.

**Supporting Note 2 | Characterization of SmC$_F$ phase for nBOE (n = 4–6).**

According to the XRD results, the *d*-spacing within the SmC$_F$ regime is according to the molecular length of **nBOE** (n = 4–6), implying that this smectic phase may possess SmA order rather than SmC. However, in the antiparallel-rubbed cell (5 μm), the SmC$_F$ phase showed a fine stripe along the rubbing direction (Figure S6). This result is unlike the typical SmA$_F$ phase [S7] in which a uniform director alignment parallel to the rubbing direction is allowed. Let us now discuss the plausible structure of the SmC$_F$ phase. Based on the X-ray diffraction pattern, this phase can be categorized as a smectic LC phase. In the case of a cell (gap: 2 μm), the strong light scattering was effectively suppressed, thereby showing the dark field due to the excellent molecular alignment along the rubbing direction (Figure S6a); however, the coexistence of the two domains was visualized by the insertion of a sensitive tint-plate (Figure S6b). In addition, upon the rotation of the stage by ± 20° within either domain, the birefringence changed to a gray color or retained its dark appearance (Figure S6c). Rotation of the analyzer produced a contrast inversion in the two domains (Figure S6d). Thus, in the two domains, the molecules within a layer are oriented at a finite angle with different signs relative to the normal direction of the layer. An optical tilt angle ($\theta_{opt}$) in the SmC$_F$ phase was determined using a microscopic method (Figure S7), resulting in $\theta_{opt}$ ~2–3°. Thus, the combined data demonstrate that the third mesophase exhibited for **nBOE** (n = 4–6) is the SmC$_F$ phase with a small tilt angle (≤ 3°).

**Supporting Note 3 | Characterization of 7BOE and 8BOE.**

In the DSC curves for **7BOE** and **8BOE** (Figure S8), a small exothermal peak ($\Delta H$ = 20–40 J mol$^{-1}$) appeared at ~175 °C, attributed to the N–N′ phase transition. In the case of **7BOE**, an additional exothermal peak emerged at 134.8 °C ($\Delta H$ = 0.11 kJ mol$^{-1}$) due to the N′–SmA phase transition. These two peaks are also observed upon heating, indicating that the three phases exhibit enantiotropic properties. Similarly, **8BOE** displayed two peaks at 175.2 °C and ~80 °C. For **8BOE**, the bottom mesophase was metastable. Figure S9 shows POM images of the planar (annealed PMMA) and parallel-rubbed cells for **7BOE** and **8BOE**. In the planar cell, a typical Schlieren texture, a broken Schlieren texture, and a blocky texture were observed in the N, N′, and SmA phases, respectively (Figure S9a). In the rubbed cell, the SmA phase exhibits a blocky striped texture, whose birefringence vanishes at the extinction position (Figure S9b). This result indicates that the molecules in a layer were aligned normal to the lamellar structure. DR and *P–E* studies failed to produce any data characteristic of the ferroelectric properties in the N, N′, and SmA phases (Figure S10). Thus, N, N′, and SmA phases possess a paraelectric property.

**Supporting Note 4 | FTIR spectra measurement.**

Figure S11 shows the computational and experimental FTIR spectra of **5BOE**. Figure S11a shows a discrepancy between the peak position ($v_{FTIR}$ or $v_{DFT}$) on the characteristic spectra, attributed to the computational results considering the gas phase. Nevertheless, no significant differences in the peak intensities or relative peak positions were observed. Thus, the respective vibration modes were characterized by calibrating the peak shifts. For example, the peak at $v_{FTIR}$ = 1630 cm$^{-1}$ ($v_{DFT}$ = 1670 cm$^{-1}$) was identified predominantly as a C–C stretching vibration of the second aromatic ring (Ar$^{II}$).

**Supporting Note 5 | Control materials: nBOE-F and nBOE-NO₂.**

Figure S12a displays the phase transition behavior for **4BOE**, **4BOE-F** and **4BOE-NO₂**. All compounds exhibited the enantiotropic $N_F$ phase with a broaden temperature window. As shown in Figure S12a and S12b, three types of compounds showed the similar trend with respect to dipole moment and $\beta$ angle as a function of n, and almost same corresponding magnitude (see also Table S1). Nevertheless, the $^{HC}N_F$ phase was absent in **4BOE-F** and **4BOE-NO₂**, indicating that an interaction with a –CN group contributes emerging the $^{HC}N_F$ phase.

**Supporting Figures (Figures S1–S30)**

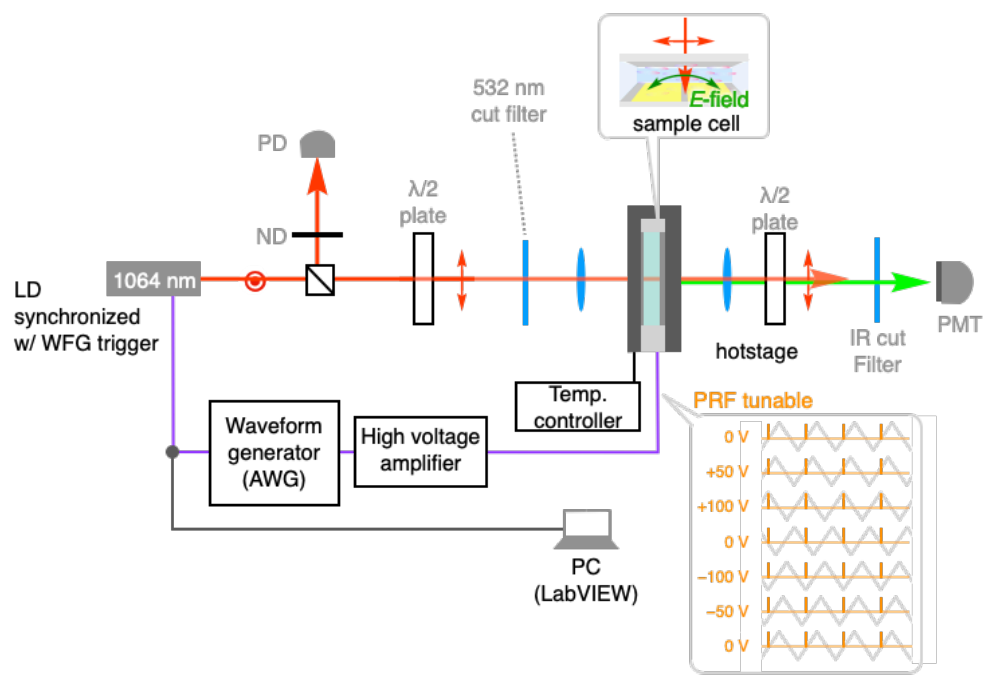

**Figure S1 Optical setup for SHG studies.** The electric field (*E*-field) was applied normally to the sample cell (thickness: 5 µm).

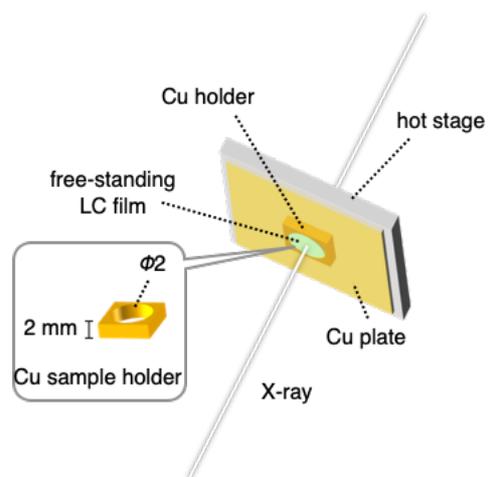

**Figure S2 Measurement setup for XRD studies.**

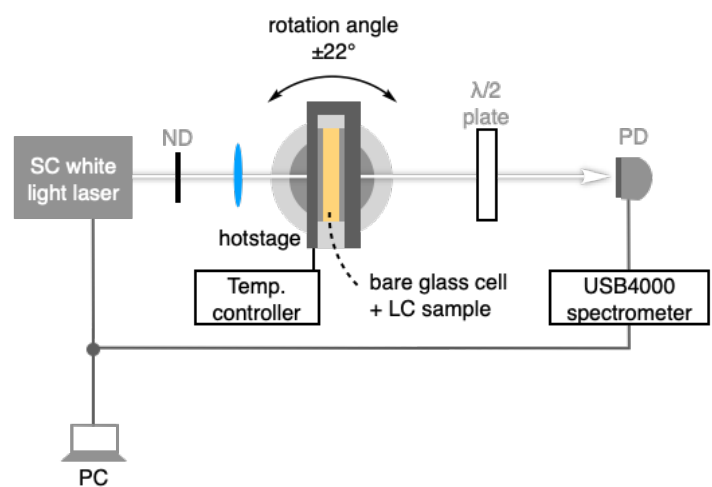

**Figure S3 Optical setup for Spectra studies.** Cell thickness: 5 µm.

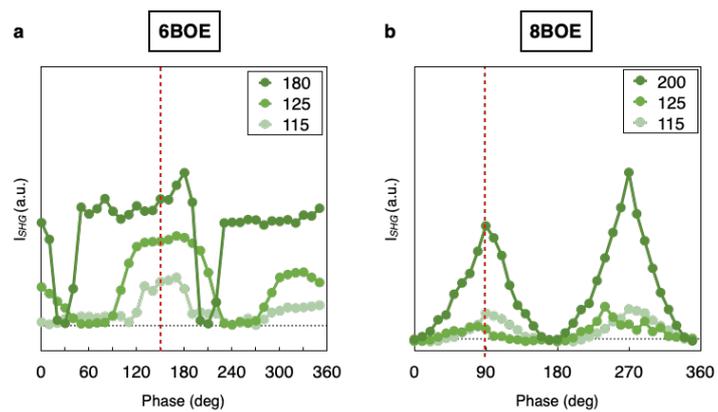

**Figure S4** SHG Intensity vs phase (related to the magnitude of *E*-field) in various temperature for **6BOE** and **8BOE**.

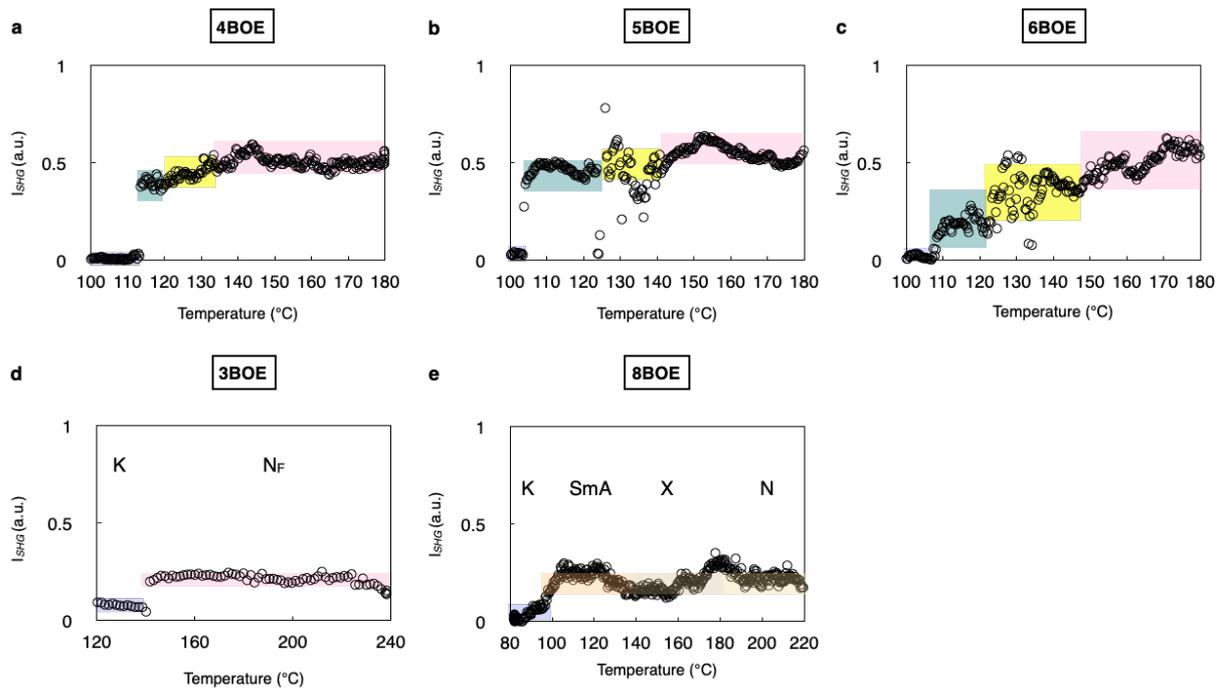

**Figure S5** SHG vs temperature for **nBOE** (n = 3–6, 8).

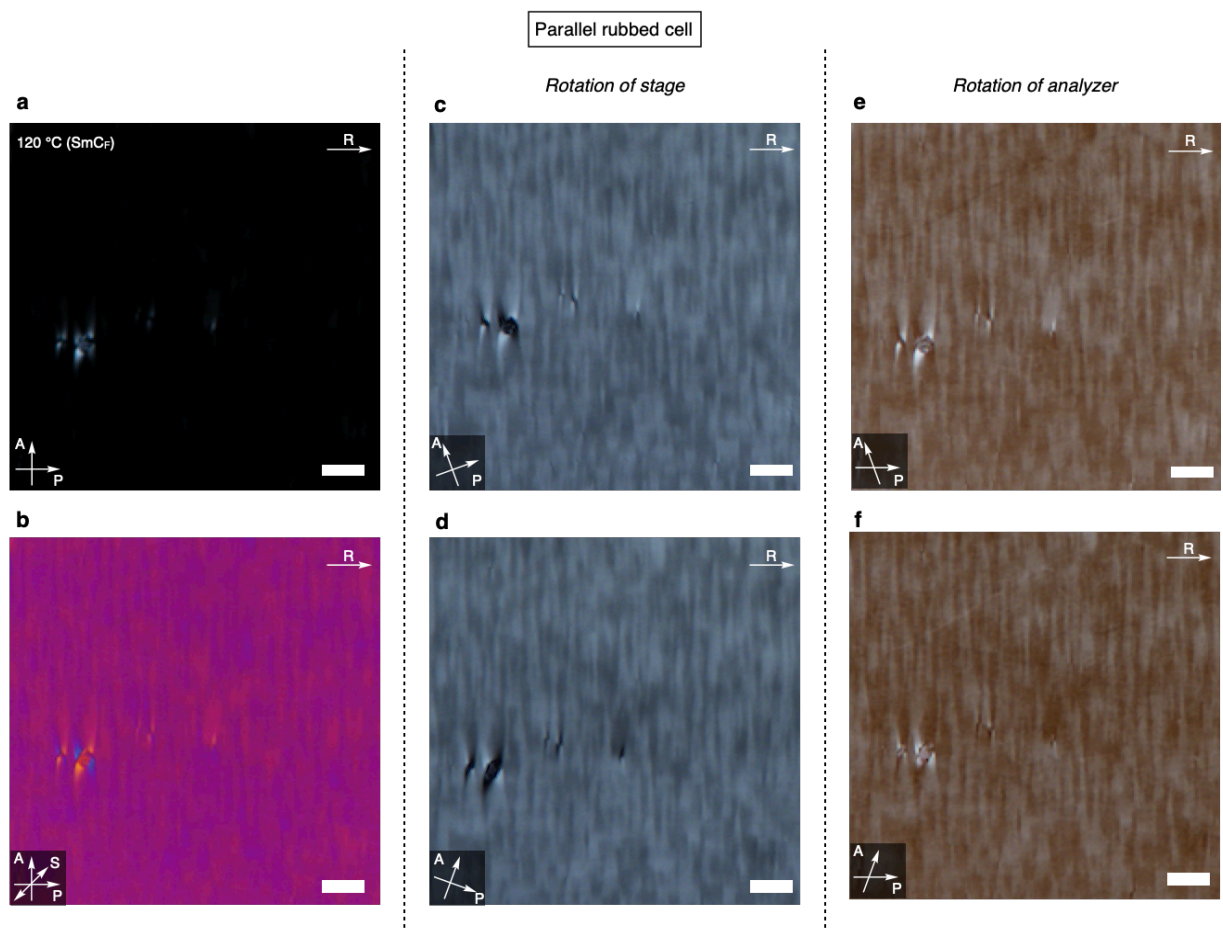

**Figure S6** POM images of the SmC$_F$ phase (110 °C) in the parallel rubbed cell (2 μm). Before (a) and after (b) insertion of the sensitive-tint plate. The rotation of stage by +20° (c) and −20° (d). The rotation of analyzer by +20° (e) and −20° (f). Scale bar: 100 μm.

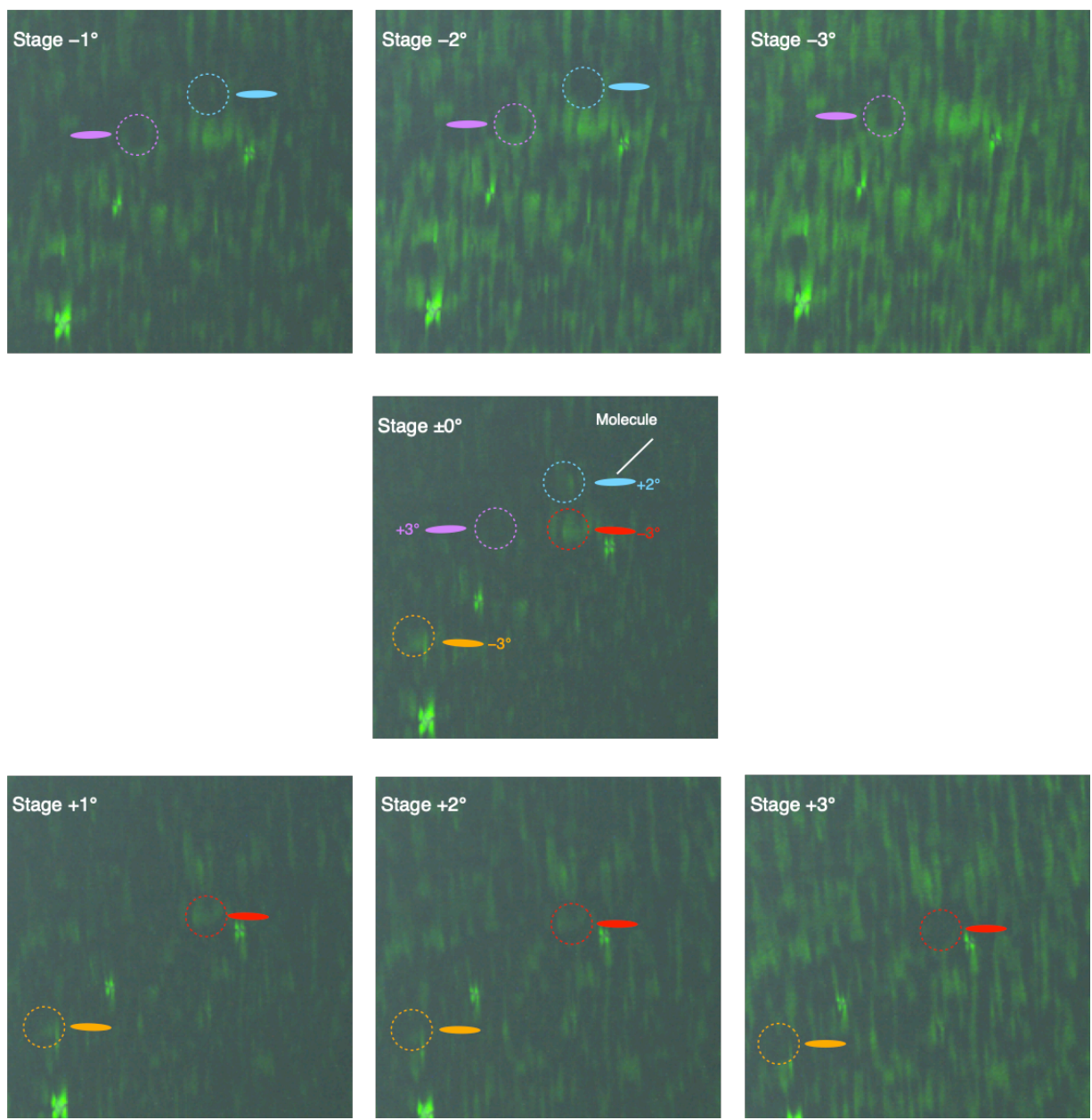

**Figure S7** Determination of an optical tilt angle in the SmC$_F$ phase (110 °C, 2 μm thicker rubbed cell) using a microscopic method (green band path filter: 532 nm).

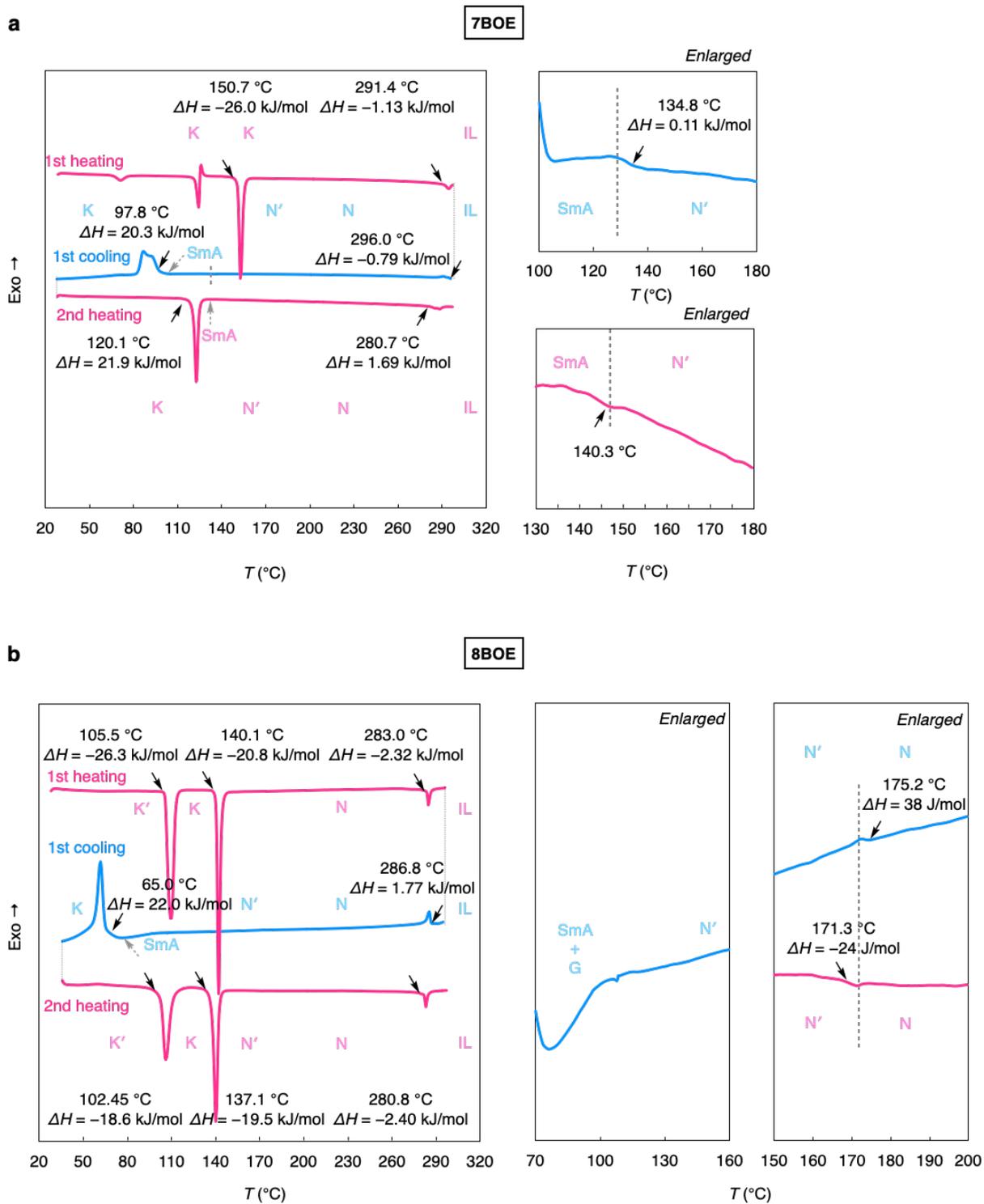

**Figure S8** DSC curves for **7BOE** (a) and **8BOE** (b). Scan rate 20 K min⁻¹.

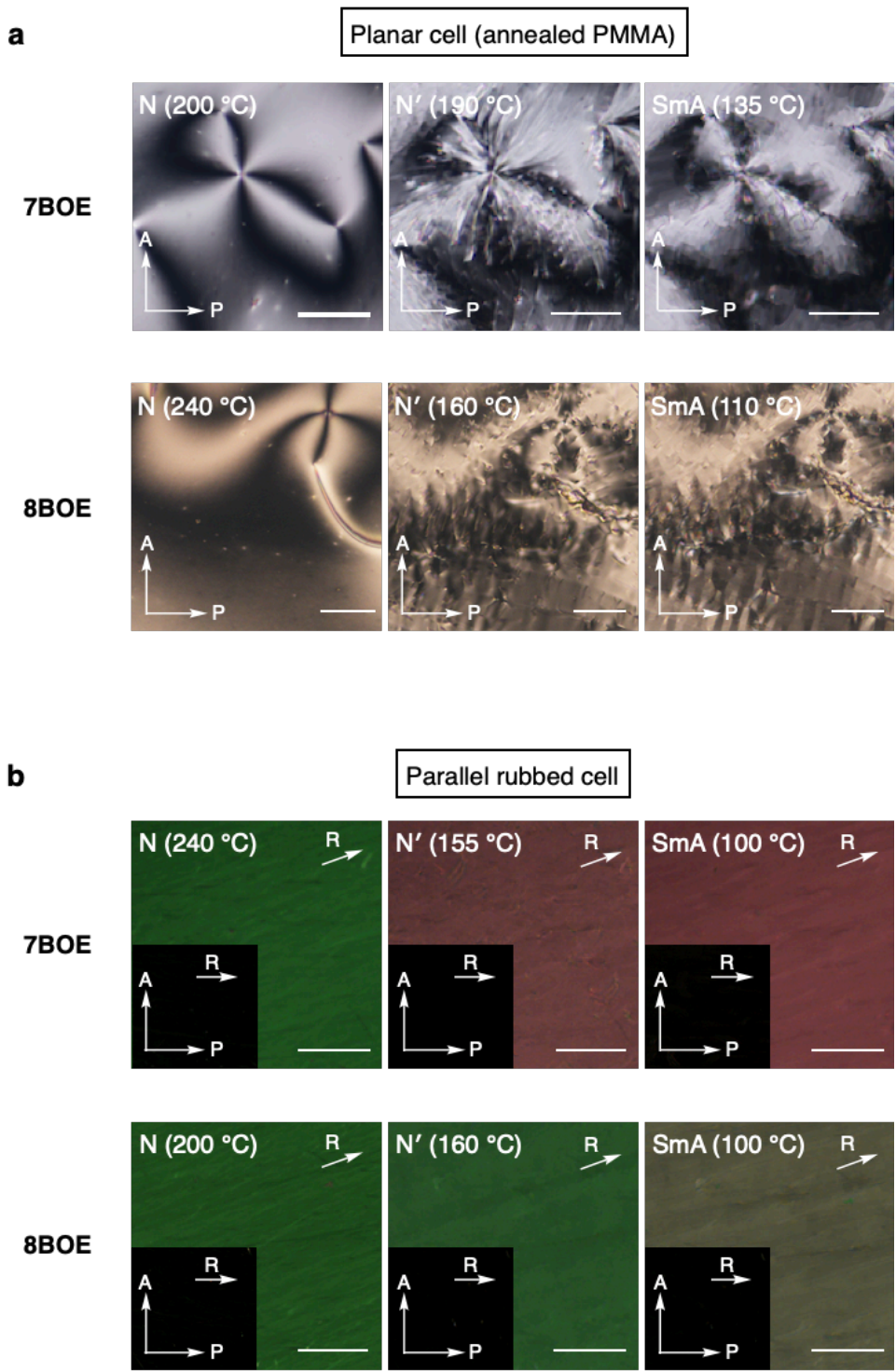

**Figure S9** Extra POM images for **7BOE** and **8BOE** in the planar (upper) and antiparallel rubbed (bottom) cells. Bar: 100 μm.

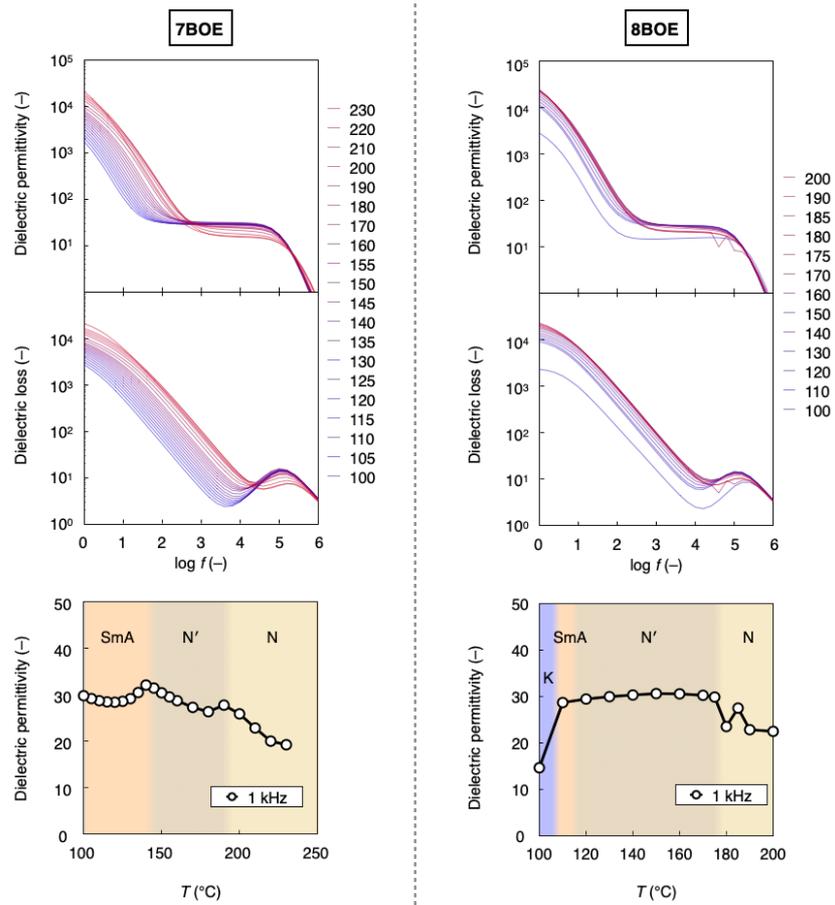

**Figure S10** Extra DR spectra for **7BOE** and **8BOE**.

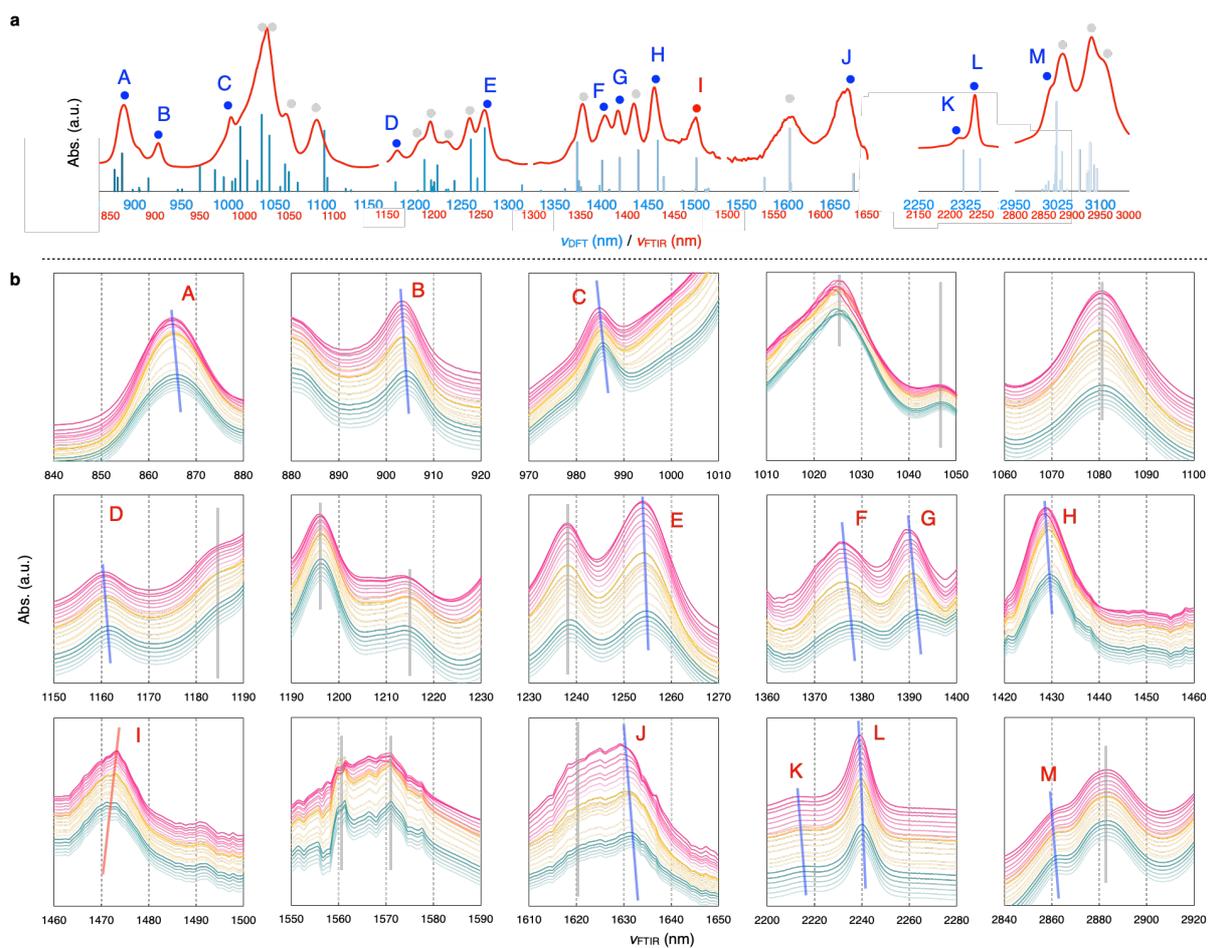

**Figure S11** (a) Computational and Experimental FTIR spectra for **5BOE**. (b) Highlighted FTIR spectra during $N_F$–$^{HC}N_F$–$SmC_F$ phase transition.

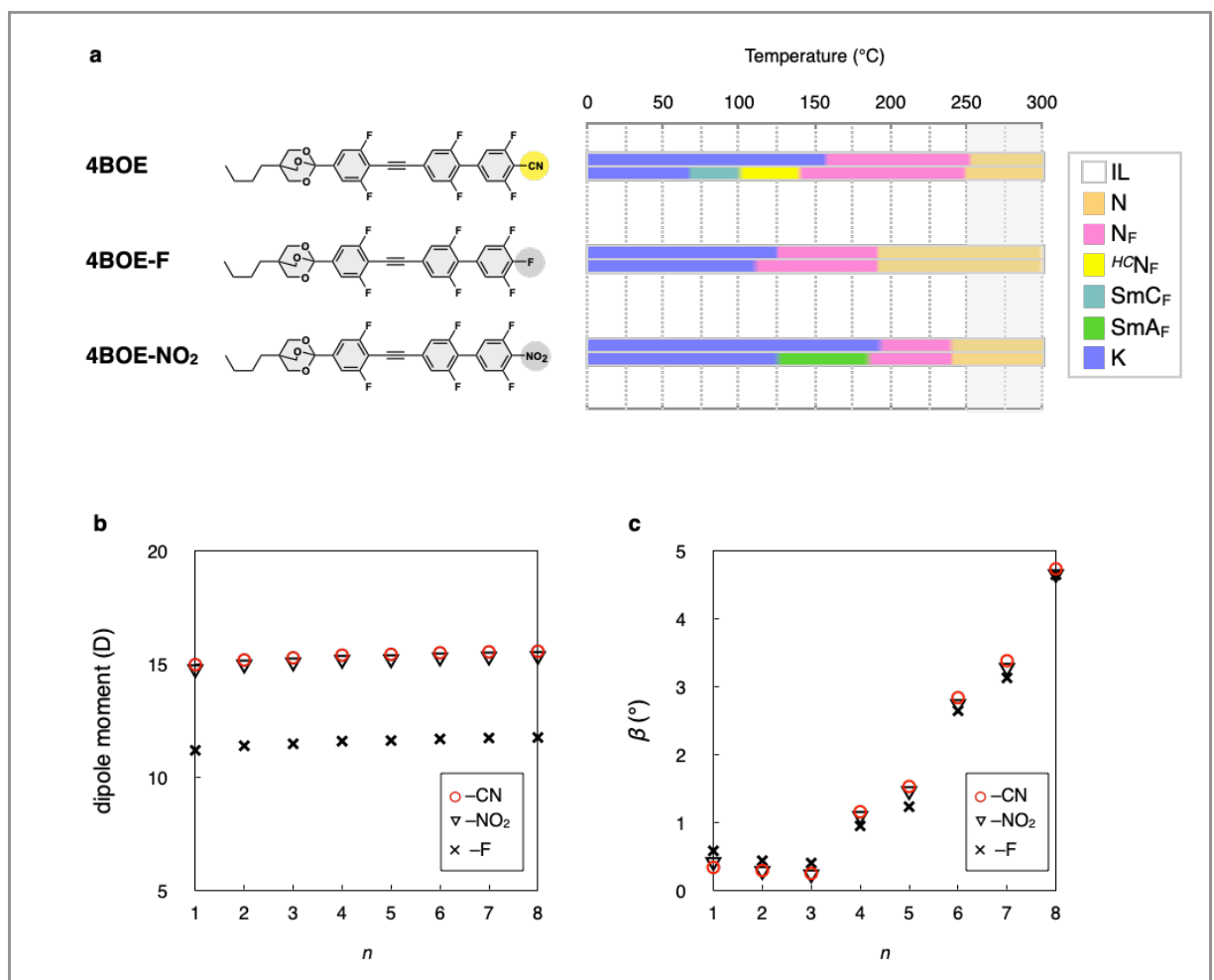

**Figure S12** Comparison of phase transition behavior (a), dipole moment (b) and $\beta$ angle (c) vs carbon number of alkyl chain (n) for **4BOE**, **4BOE-F** and **4BOE-NO₂**.

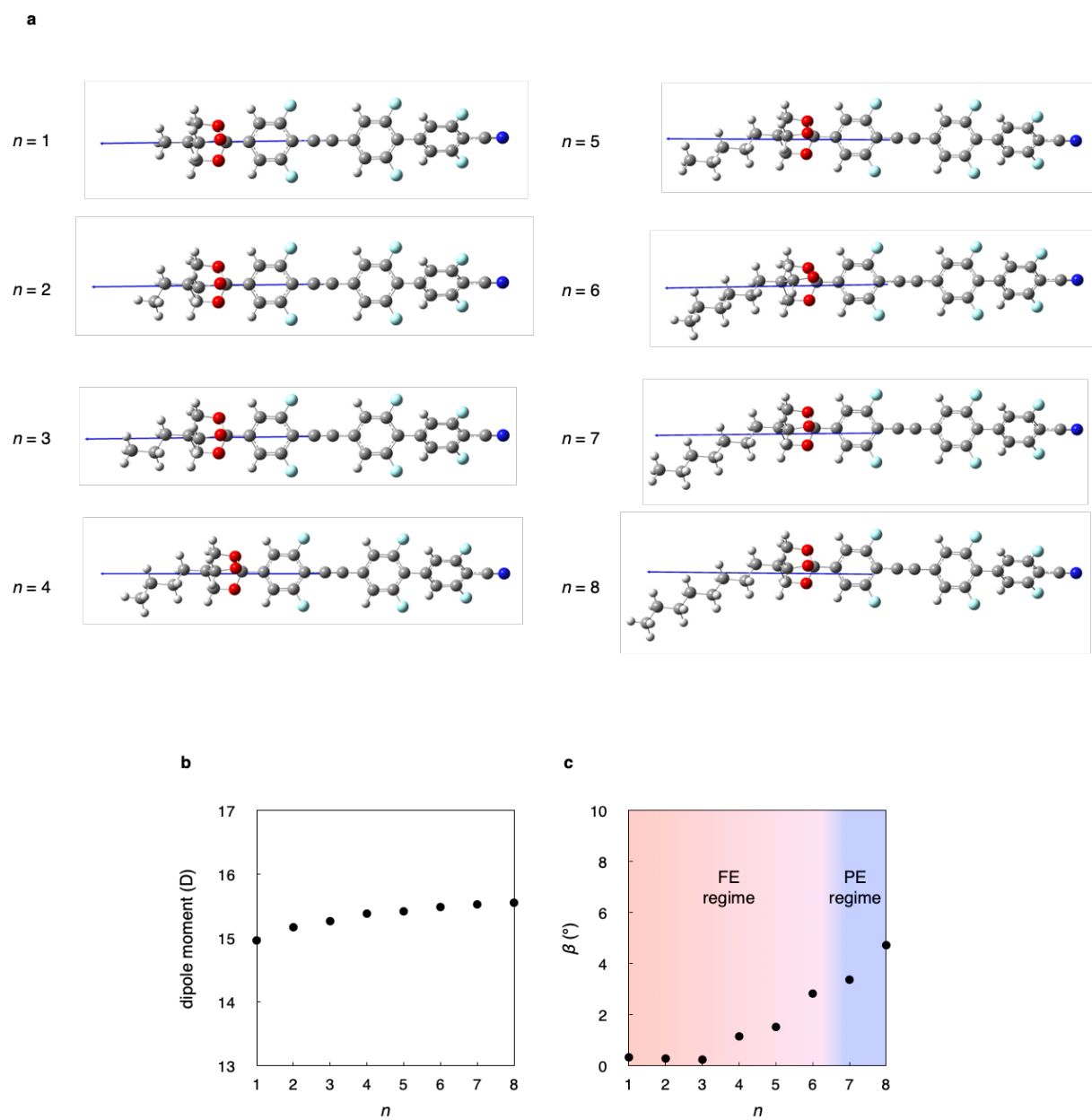

**Figure S13** (a) Optimized structures of **nBOE** (n = 1–8). Dipole moment (a) and $\beta$ angle (b) vs carbon number of alkyl chain (n).

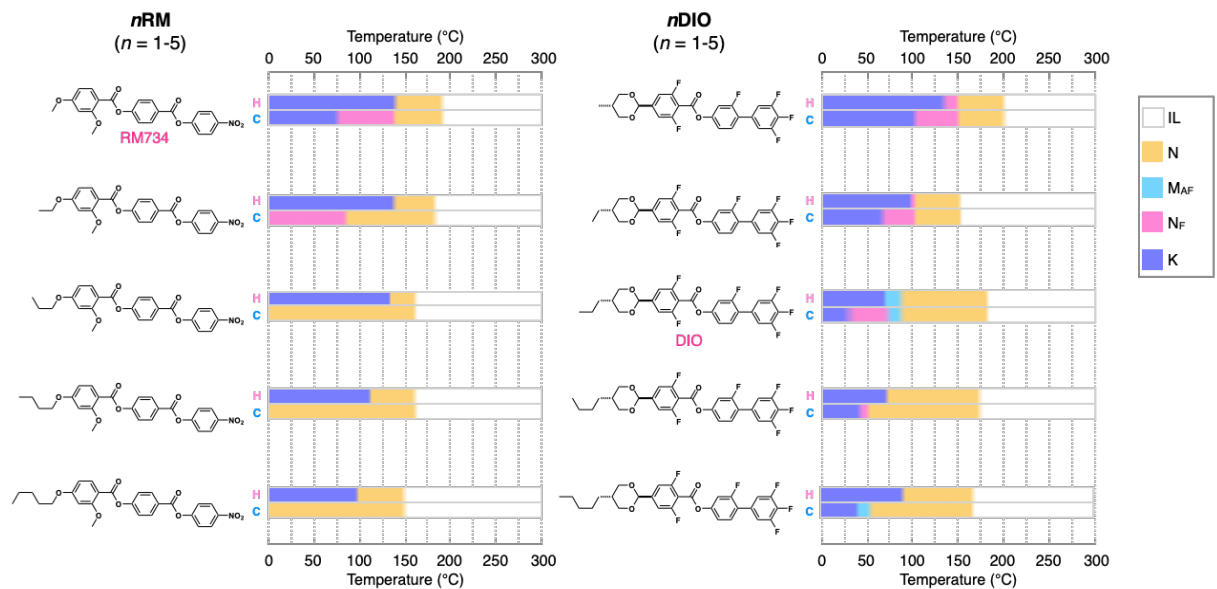

**Figure S14** Phase transition behavior for **nRM** (n = 1–5) and **nDIO** (n = 1–5). Data of **nRM** (n = 1–5) were extracted from ref. S8. For data of **nDIO** (n = 2,3,4) were extracted from ref. S9, S10 and S11, respectively. The phase transition temperature for **nDIO** (n = 1,5), which were synthesized in our laboratory, was decided by DSC studies.

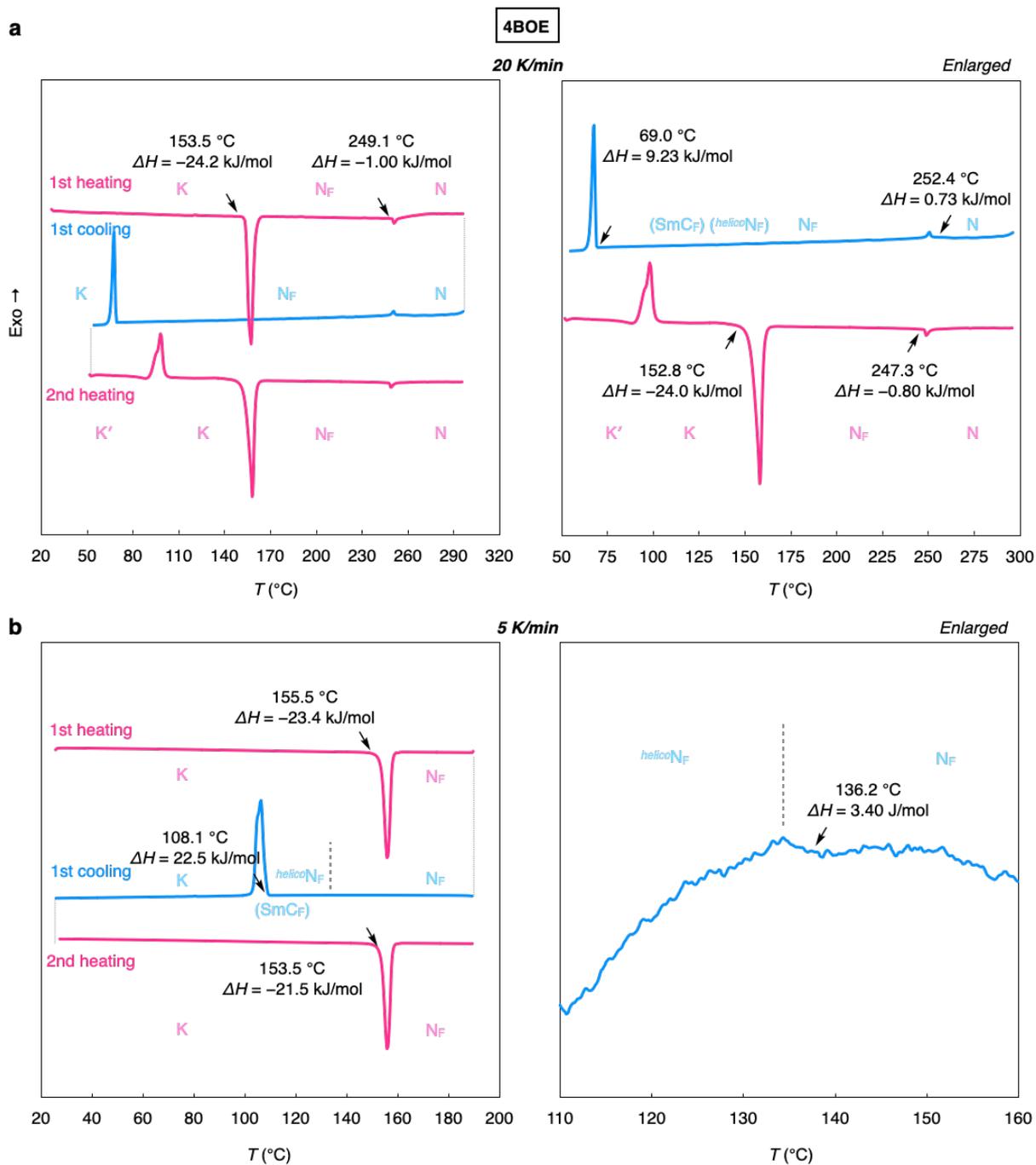

**Figure S15** DSC curves for **4BOE**. Scan rate 20 K min⁻¹ (a) and 5 K min⁻¹ (b).

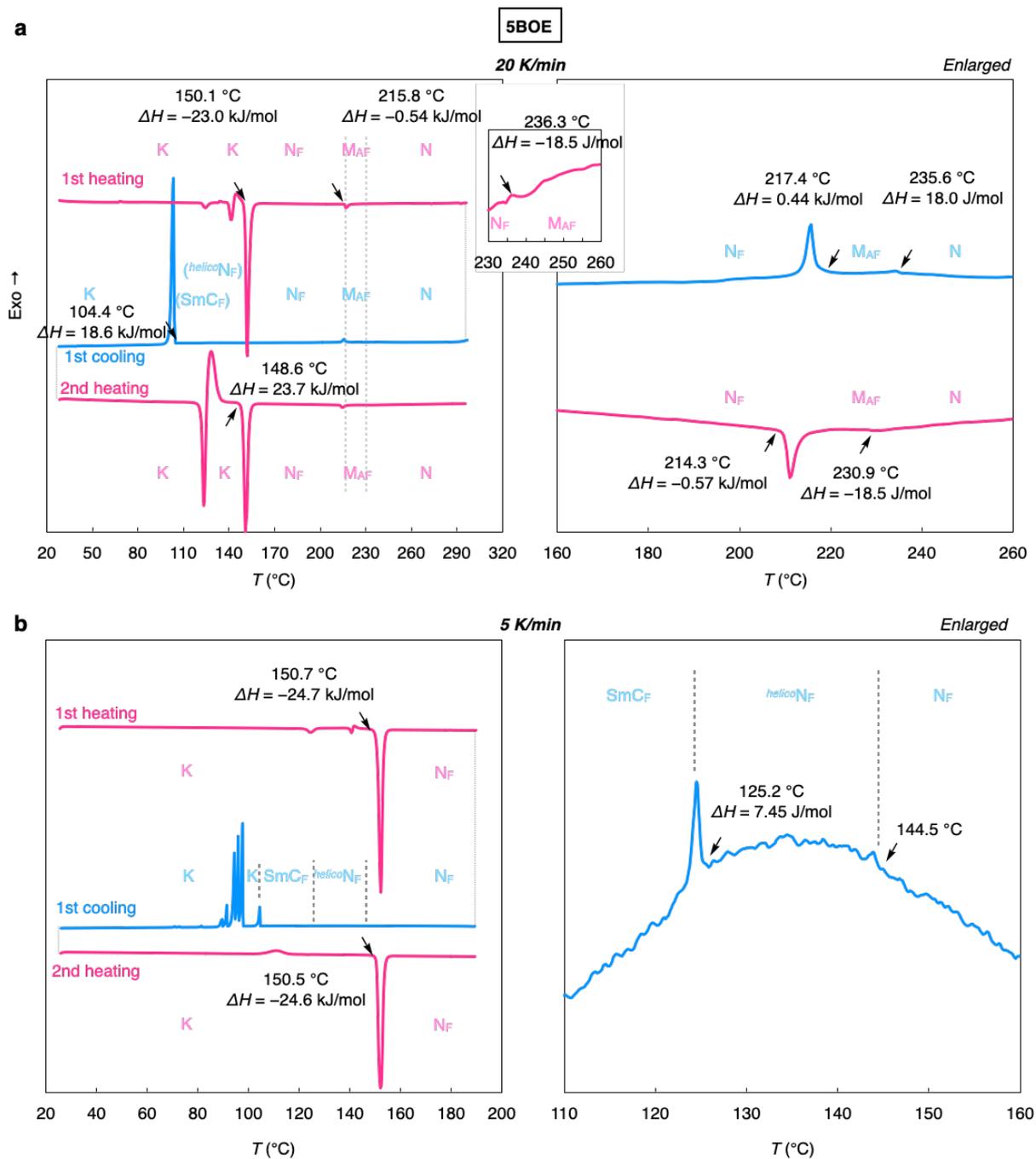

**Figure S16** DSC curves for **5BOE**. Scan rate 20 K min⁻¹ (a) and 5 K min⁻¹ (b).

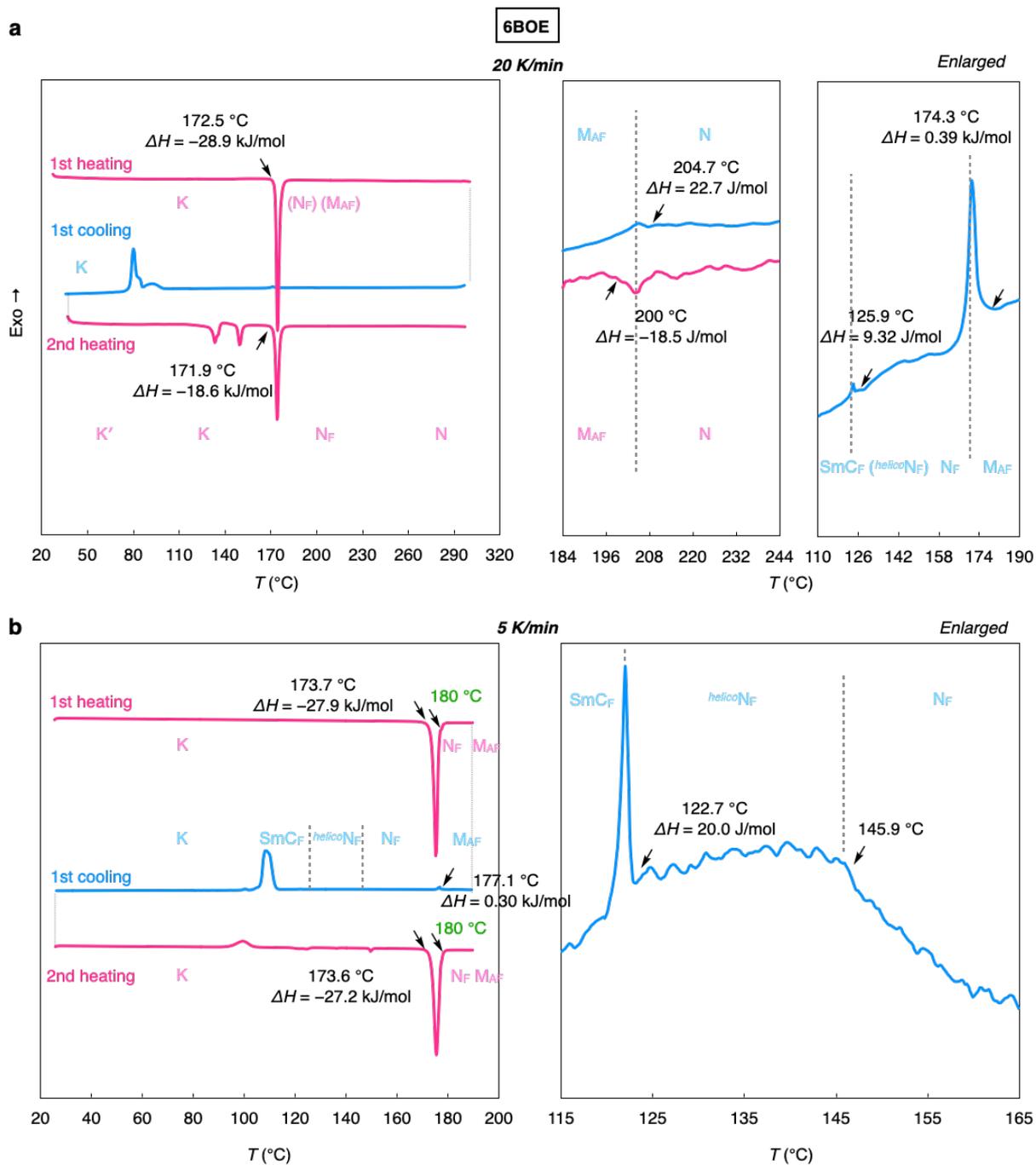

**Figure S17** DSC curves for **6BOE**. Scan rate 20 K min⁻¹ (a) and 5 K min⁻¹ (b).

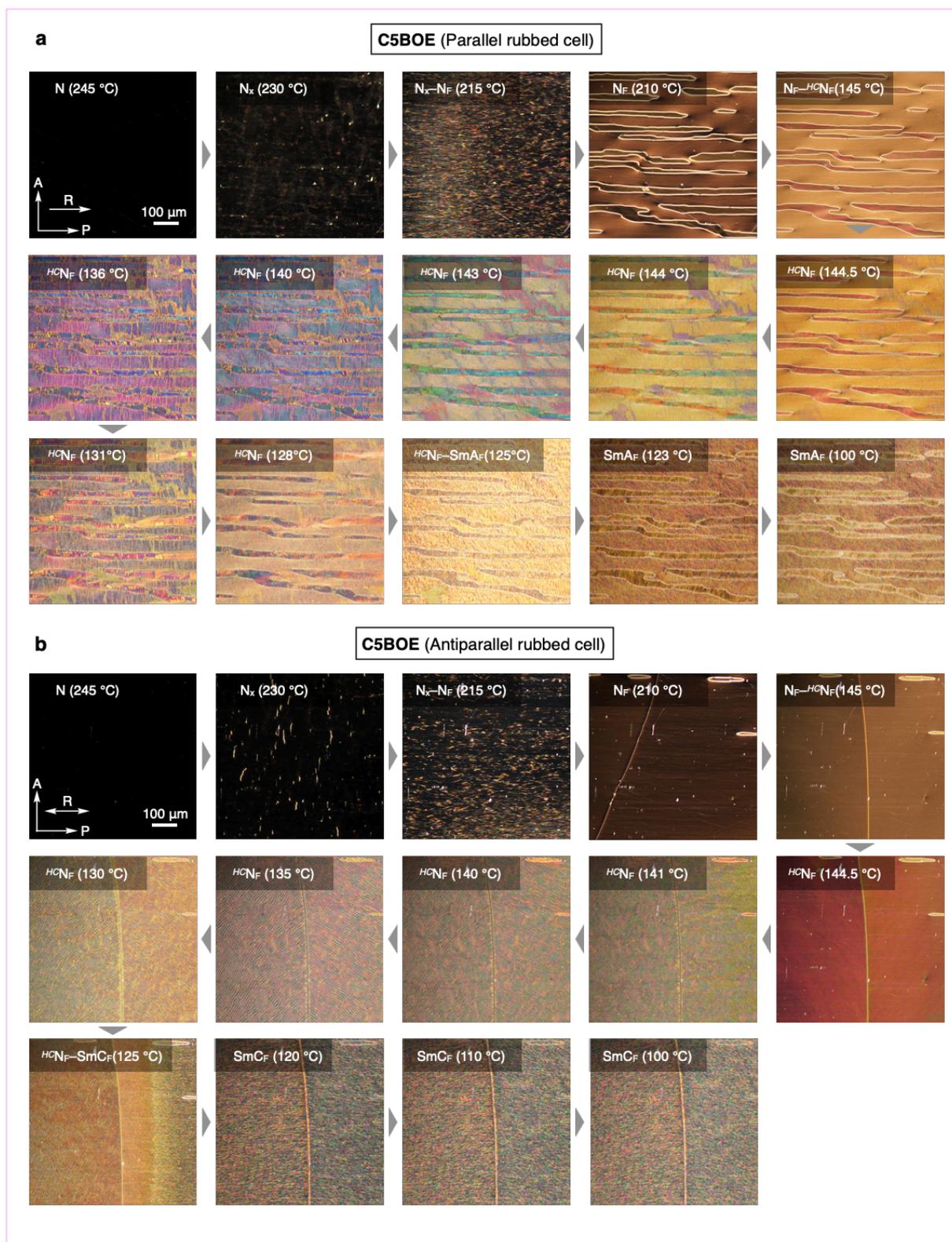

**Figure S18** Extra POM images for **5BOE** in the parallel- (a) and antiparallel- (b) cells. Scale bar: 100 μm.

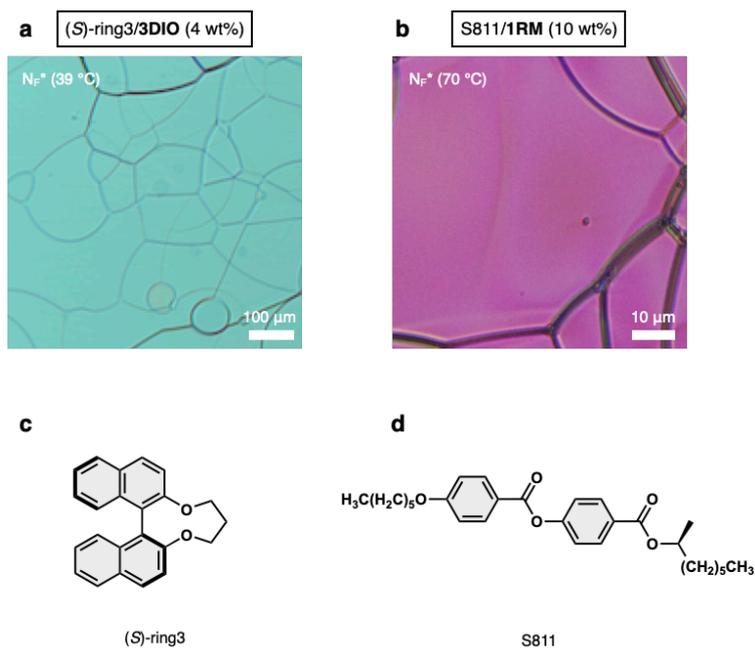

**Figure S19** POM images of the ferroelectric cholesteric (N$_F$*) phase in two cases: (a) (*S*)-ring/**3DIO** (4 wt%), (b) S811/**1RM** (10 wt%). Chemical structure of the corresponding chiral dopant are shown in the panel (c) and (d).

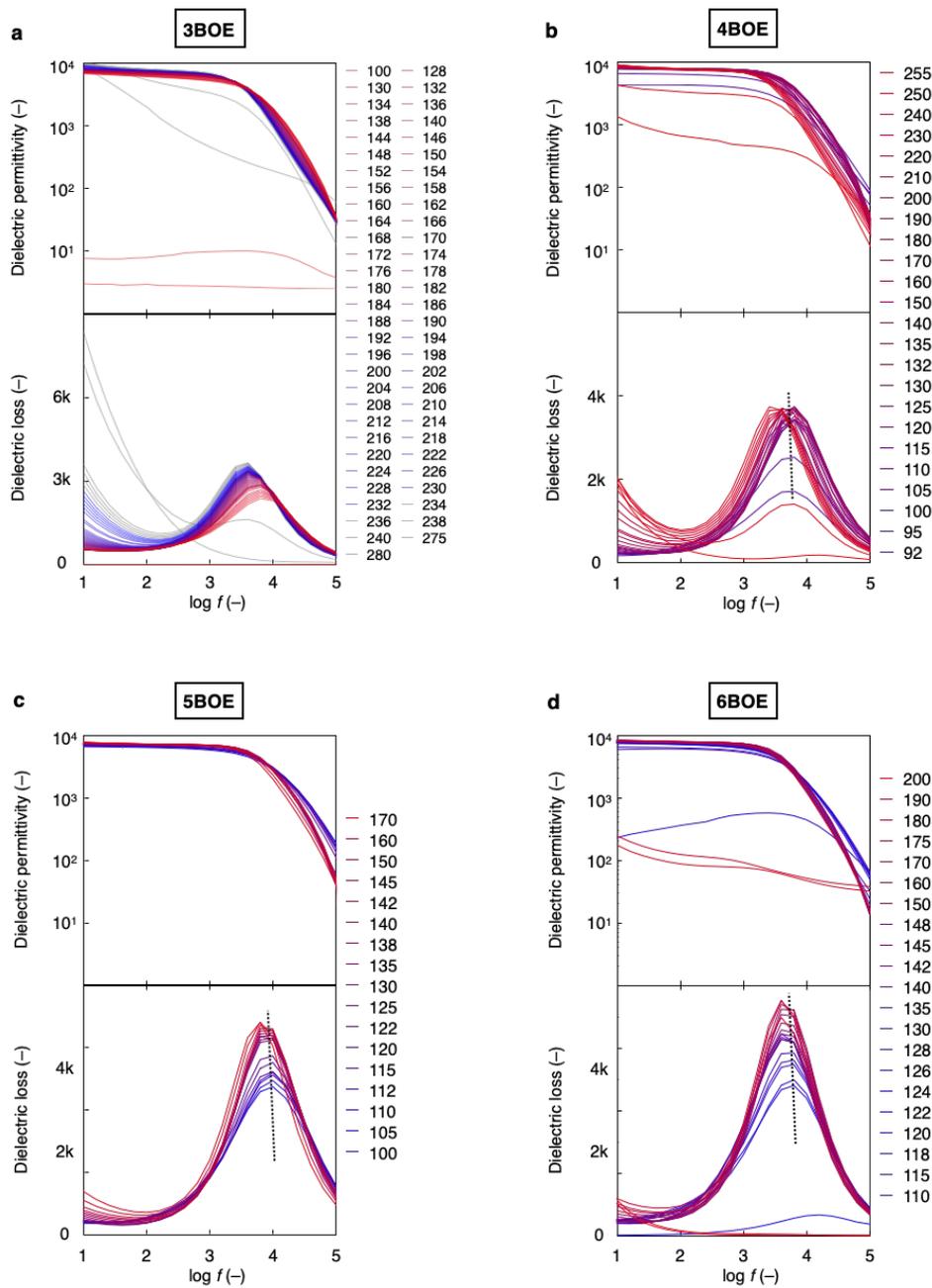

**Figure S20** Extra DR spectra for **nBOE** (n = 3–6).

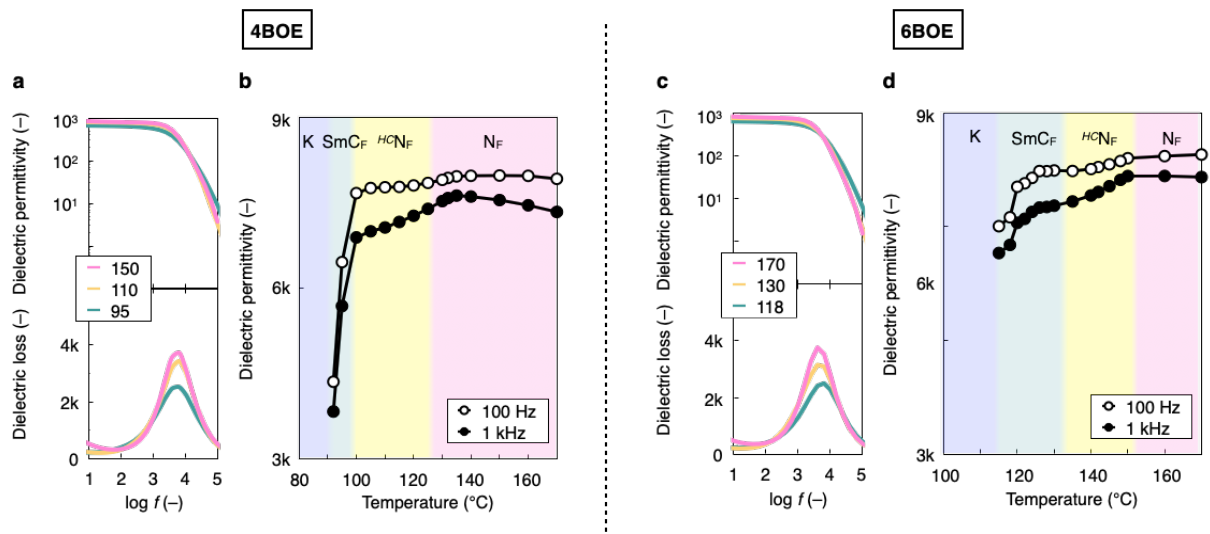

**Figure S21** DR spectra and temperature-dependent dielectric permittivity for **4BOE** and **6BOE**.

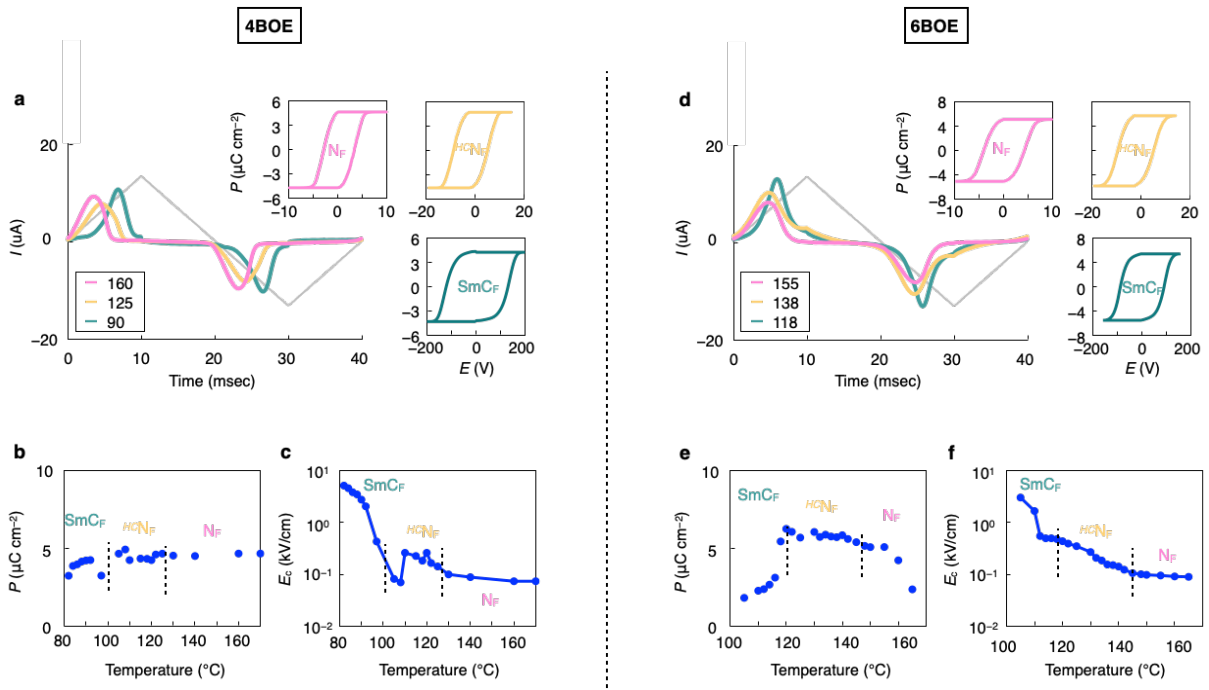

**Figure S22** Polarization behavior for **4BOE** and **6BOE**. (a,d) *I* vs time. Insets represent the corresponding *P-E* hysteresis loop. (b,e) *P* vs temperature. (c,f) $E_c$ vs temperature.

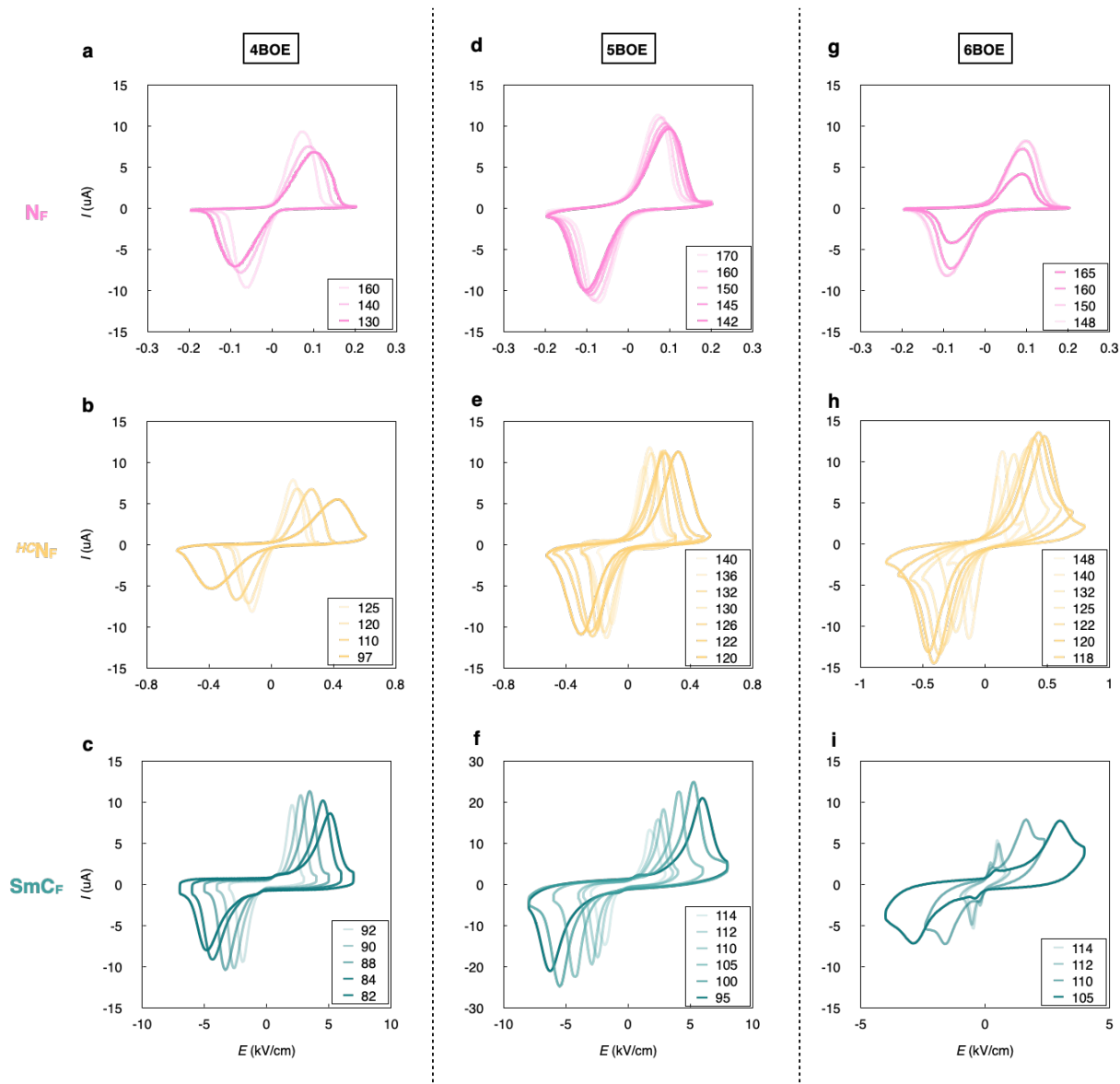

**Figure S23** Extra polarization reversal current data for **nBOE** (n = 4–6).

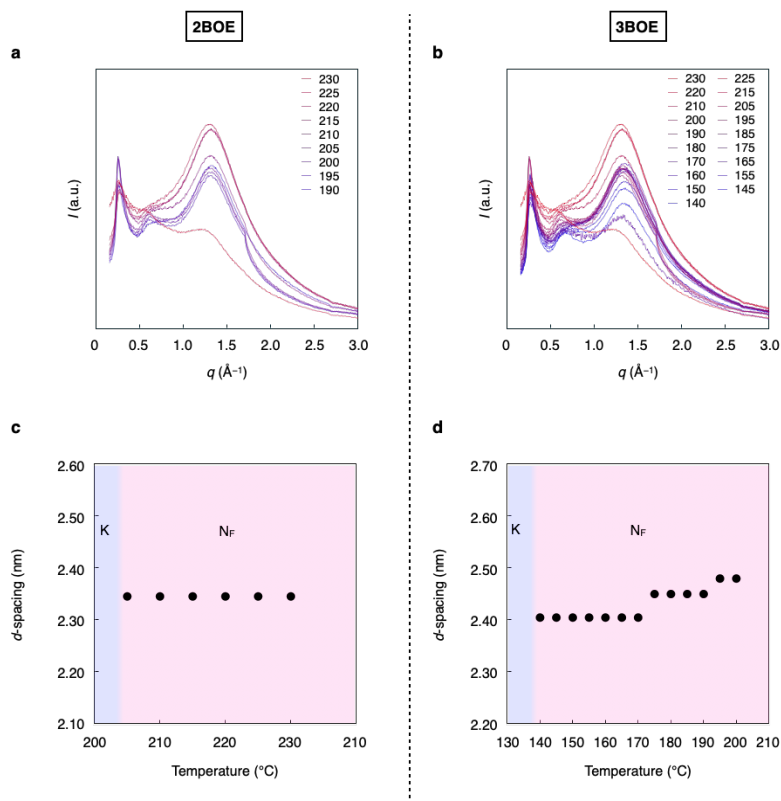

**Figure S24** Temperature dependent 1D XRD pattern and *d*-spacing for **2BOE** (a,c) and **3BOE** (b,d).

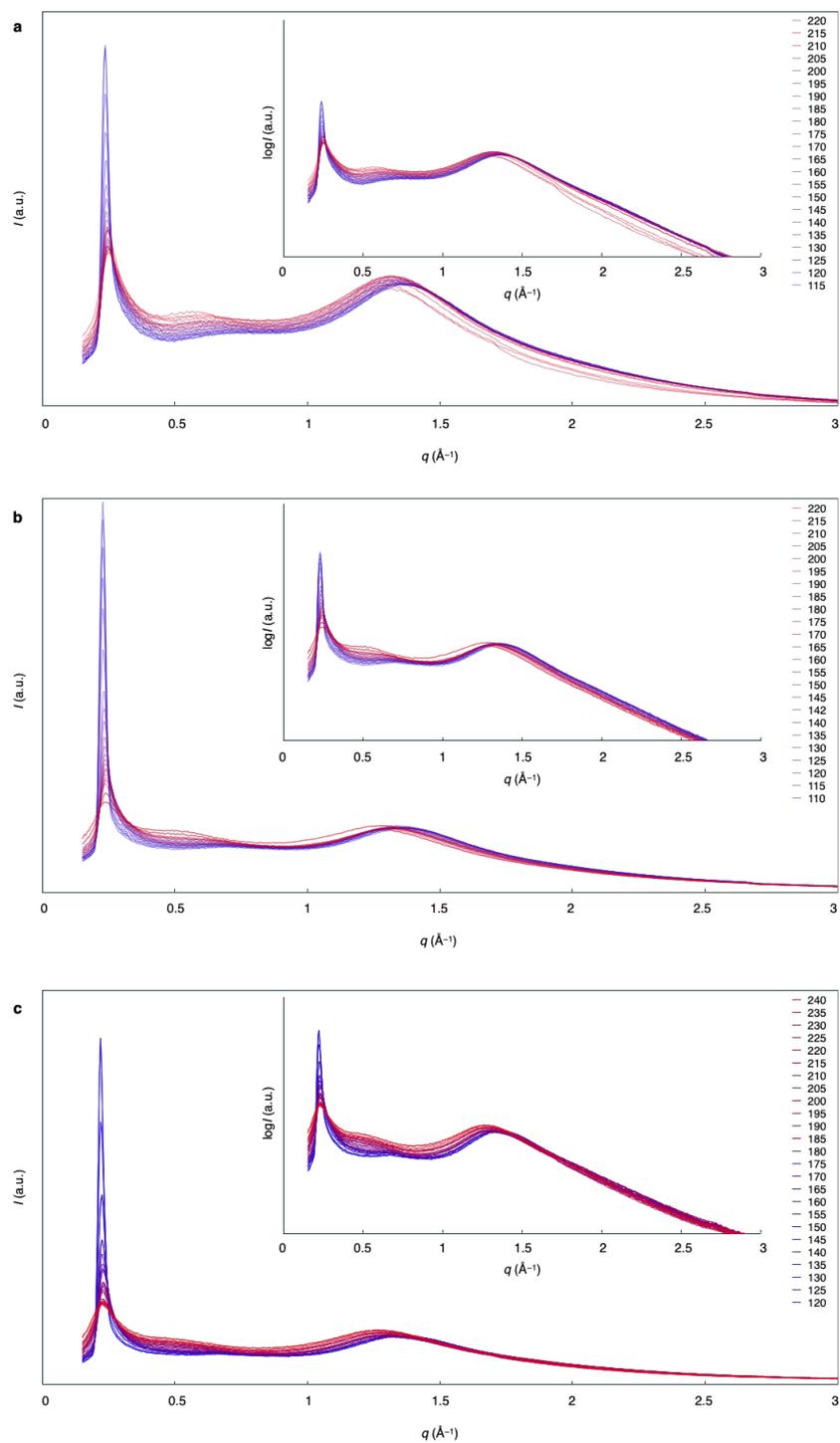

**Figure S25** Temperature-dependent 1D XRD pattern for **4BOE** (a), **5BOE** (b) and **6BOE** (c).

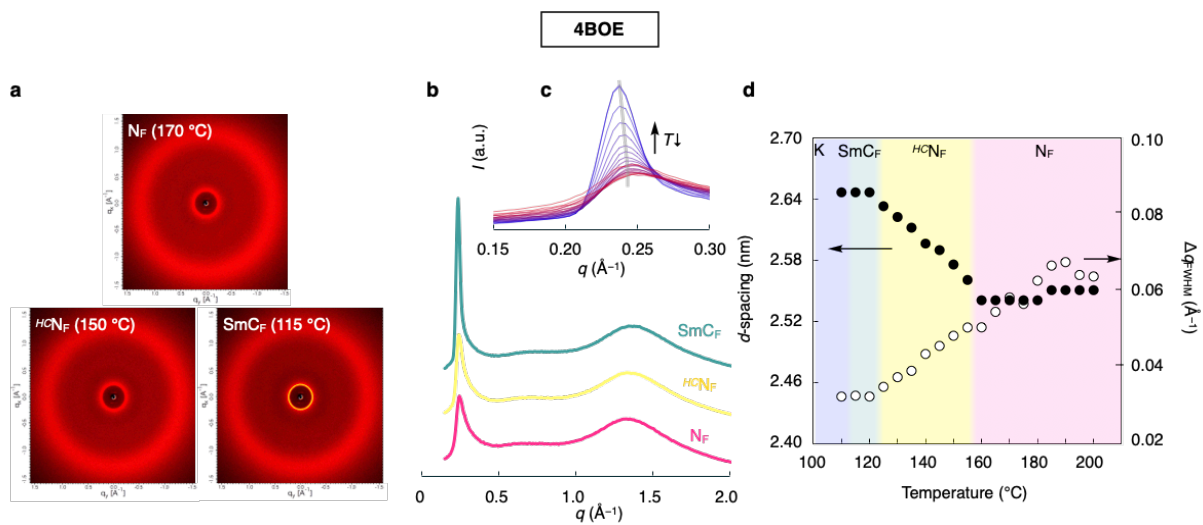

**Figure S26** XRD studies for **4BOE**. 2D XRD (a) and 1D XRD (b,c) pattern in various phases. (d) $d$-spacing and FWHM vs Temperature. For the panel (b), the recorded temperature in $N_F$, $^{HC}N_F$ and $SmC_F$ phases are 170, 150 and 115 °C, respectively.

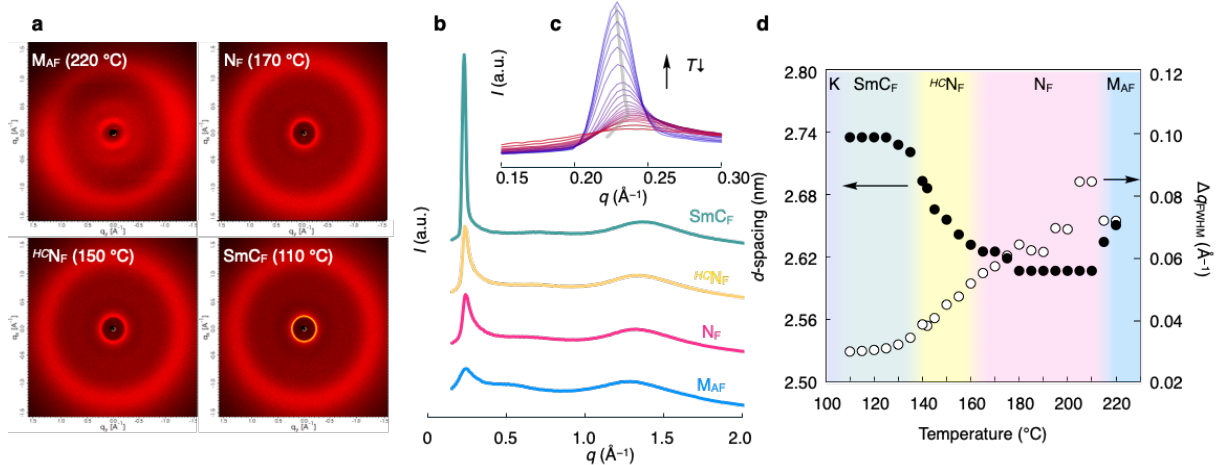

**5BOE**

**Figure S27** XRD studies for **5BOE**. 2D XRD (a) and 1D XRD (b,c) pattern in various phases. (d) *d*-spacing and FWHM vs Temperature. For the panel (b), the recorded temperature in $M_{AF}$, $N_F$, $^{HC}N_F$ and $SmC_F$ phases are 220, 170, 150 and 110 °C, respectively.

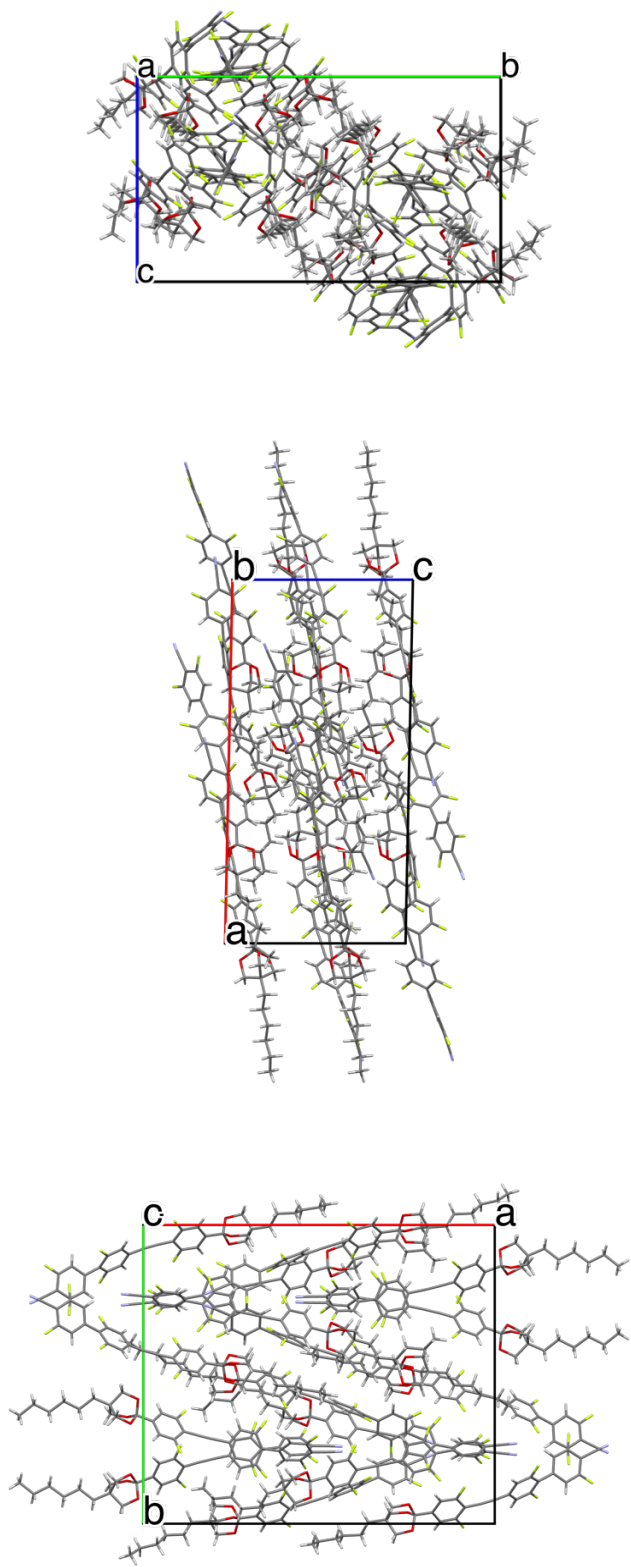

**Figure S28 SC-XRD data** viewed along the a-, b-, and c-axis for **6BOE**: crystal twinning, crystal system: monoclinic, space group: $P2_{1/c}$, cell length: **a** = 29.5406(2) Å, **b** = 25.3902(2) Å, **c** = 14.49345(13) Å, cell angle: **α** = 90°, **β** = 91.1515(7)°, **γ** = 90°, cell volume: **V** = 10868.5 Å$^3$, **Z** = 16, **Z′** = 4, R-factor = 7.71%.

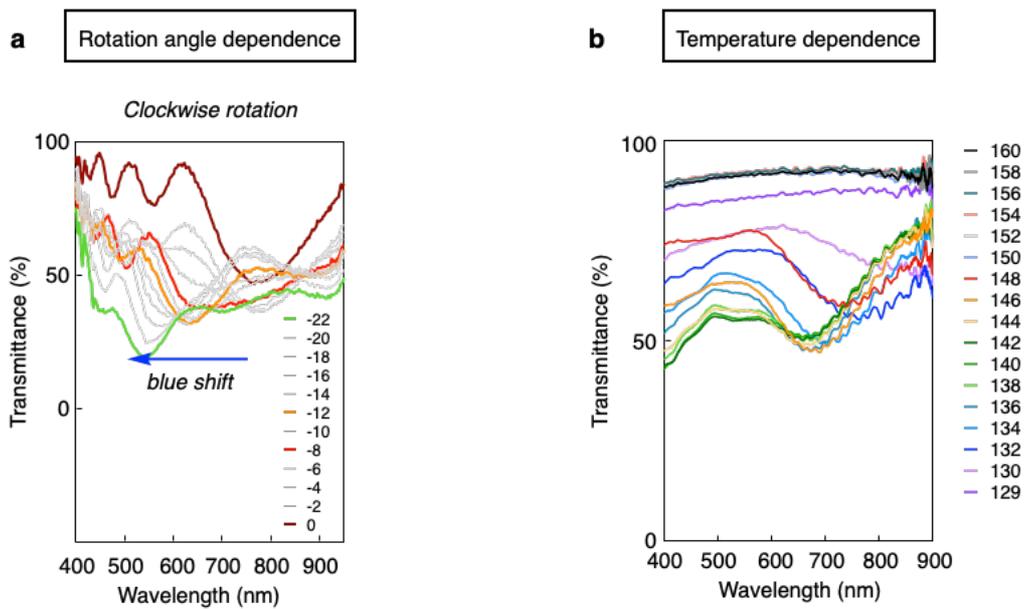

**Figure S29 Extra spectra data for 5BOE.** (a) Rotation angle dependence ($^{HC}$N$_F$ phase at 145 °C), (b) temperature dependence (rate: 1 K min$^{-1}$).

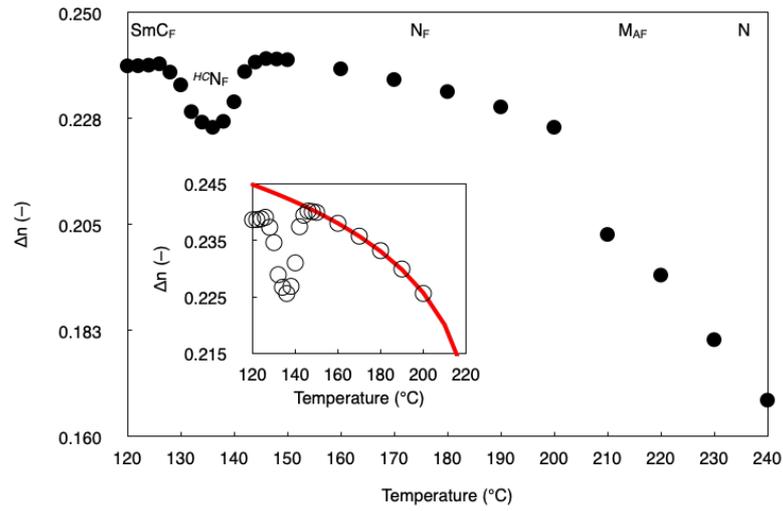

**Figure S30 Temperature dependence of 5BOE.** Inset: fitting curve calculated within $N_F$ regime. The $\Delta n(N_F)$ was extrapolated to the lower temperature range by supposing a power-law temperature dependence: $\Delta n(N_F) = \Delta n_0(T_c - T)^\gamma$, where $\Delta n0$, $T_c$, and $\gamma$ are the fitting parameters.

## Supporting Tables (Tables S1 and S2)

**Table S1** The molecular parameters of energy-minimized conformations calculated by MM2/DFT for **BOE** variants.

| Entry | X | Y | Z | $\mu$ (D) | $\beta$ (deg)† |
|---|---|---|---|---|---|
| **1BOE** | 14.967 | 0.037 | -0.080 | 14.967 | 0.339 |
| **2BOE** | 15.173 | 0.061 | -0.048 | 15.173 | 0.293 |
| **3BOE** | 15.269 | 0.062 | -0.023 | 15.269 | 0.249 |
| **4BOE** | -15.385 | -0.254 | 0.179 | 15.388 | 1.158 |
| **5BOE** | -15.417 | -0.320 | 0.259 | 15.423 | 1.529 |
| **6BOE** | -15.482 | -0.642 | 0.420 | 15.491 | 2.836 |
| **7BOE** | -15.472 | -0.808 | 0.427 | 15.532 | 3.379 |
| **8BOE** | -15.505 | -1.155 | 0.560 | 15.558 | 4.732 |
| **1BOE-NO₂** | 14.678 | 0.010 | -0.099 | 14.678 | 0.389 |
| **2BOE-NO₂** | 14.885 | 0.028 | -0.061 | 14.885 | 0.257 |
| **3BOE-NO₂** | 14.978 | 0.036 | -0.041 | 14.978 | 0.207 |
| **4BOE-NO₂** | 15.091 | 0.266 | 0.090 | 15.093 | 1.066 |
| **5BOE-NO₂** | 15.116 | 0.343 | 0.158 | 15.121 | 1.432 |
| **6BOE-NO₂** | -15.191 | -0.665 | 0.275 | 15.197 | 2.713 |
| **7BOE-NO₂** | -15.180 | -0.807 | 0.301 | 15.238 | 3.247 |
| **8BOE-NO₂** | -15.213 | -1.179 | 0.352 | 15.263 | 4.624 |
| **1BOE-F** | 11.184 | 0.011 | -0.113 | 11.184 | 0.582 |
| **2BOE-F** | 11.385 | 0.004 | -0.087 | 11.385 | 0.437 |
| **3BOE-F** | 11.473 | -0.003 | -0.081 | 11.473 | 0.402 |
| **4BOE-F** | -11.587 | -0.190 | 0.027 | 11.589 | 0.949 |
| **5BOE-F** | -11.613 | -0.239 | 0.073 | 11.615 | 1.232 |
| **6BOE-F** | -11.683 | -0.506 | 0.186 | 11.687 | 2.644 |
| **7BOE-F** | -11.675 | -0.605 | 0.204 | 11.729 | 3.128 |
| **8BOE-F** | -11.712 | -0.915 | 0.262 | 11.751 | 4.645 |

† an angle between the permanent dipole moment ($\mu$) and long molecular axis.

**Table S2** The polarization density and estimated polar order parameter <P1> for **nBOE**.

| Entry | $P_s$ (D) | <P1>[††] |
|:---:|:---:|:---:|
| **1BOE** | 4.6[†] | 0.609 |
| **2BOE** | 5.8[†] | 0.778 |
| **3BOE** | 6.5 | 0.890 |
| **4BOE** | 4.7 | 0.649 |
| **5BOE** | 6.1 | 0.869 |
| **6BOE** | 5.2 | 0.756 |

† extracted from S2, †† density ($\rho$) was set to be 1.3 g cm$^{-3}$

**Supporting References**